\documentclass[english]{article}
\usepackage[T1]{fontenc}
\usepackage[latin9]{inputenc}
\pdfoutput=1
\usepackage{refcount}
\usepackage[letterpaper]{geometry}
\usepackage{array}
\usepackage{verbatim}
\usepackage{amsmath}
\usepackage{amssymb}
\usepackage{graphicx}
\usepackage{color}
\usepackage{esint}
\usepackage[numbers]{natbib}

\usepackage{color}
\usepackage[makeroom]{cancel}
\usepackage[normalem]{ulem}
\definecolor{pkcolor}{rgb}{0,0.1,0.7}
\definecolor{ascolor}{rgb}{1,0,1}
\definecolor{mscolor}{rgb}{1,0,0}

\newcommand\pkout{\marginpar{\color{pkcolor}$\clubsuit$}\bgroup\markoverwith{\color{pkcolor}{\rule[0.4ex]{2pt}{0.8pt}}}\ULon}
\newcommand\asout{\marginpar{\color{ascolor}$\heartsuit$}\bgroup\markoverwith{\color{ascolor}{\rule[0.4ex]{2pt}{0.8pt}}}\ULon}
\newcommand\msout{\marginpar{\color{mscolor}$\diamondsuit$}\bgroup\markoverwith{\color{mscolor}{\rule[0.4ex]{2pt}{0.8pt}}}\ULon}

\newcommand\myref{\refstepcounter{equation}\theequation}
\newcommand{\refmyref}[1]{\newcounter{#1}\setcounter{#1}{\theequation}}

\begin{document}

\title{Off-shell amplitudes as boundary integrals of analytically continued Wilson line slope}

\date{}

\author{P.~Kotko$^{1}$, M.~Serino$^{2}$, A.M.~Sta\'sto$^{1}$ \vspace*{0.2cm}\\\\
$^{1}$  {\it \small  Department of Physics, 
 The Pennsylvania State University}\\ {\it \small  University Park, PA 16802, United States}\\\\
$^{2}$ {\it \small The Henryk Niewodnicza\'nski Institute of Nuclear Physics}\\ {\it \small   Polish Academy of Sciences, Krak\'ow, Poland}
}

\maketitle

\begin{abstract}
One of the methods to calculate tree-level multi-gluon scattering
amplitudes is to use the Berends-Giele recursion relation involving 
off-shell currents or off-shell amplitudes, if working in the light cone gauge.
As shown in recent works using the light-front perturbation theory, solutions to these recursions
naturally collapse into gauge invariant and gauge-dependent components,
at least for some helicity configurations. In this work, we show that
such  structure is helicity independent and emerges from  analytic
properties of matrix elements of Wilson line operators, where
the slope of the straight gauge path is shifted in a certain complex direction. 
This is similar to the procedure leading to the Britto-Cachazo-Feng-Witten (BCFW) recursion, 
however we apply a complex shift to the Wilson line slope instead of the external momenta. 
While in the original BCFW procedure the boundary integrals over the complex shift vanish for certain deformations, 
here they are non-zero and are equal to the off-shell amplitudes. 
The main result can thus be summarized as follows: 
we derive a decomposition of a helicity-fixed off-shell current into gauge invariant component given 
by a matrix element of a straight Wilson line plus a reminder given by a sum of products of gauge invariant and gauge dependent quantities. 
We give several examples realizing this relation, including the five-point next-to-MHV helicity configuration.
\end{abstract}

\section{Introduction}
\label{sec:Intro}

In applications of perturbative Quantum Chromodynamics one faces the
challenge of calculating parton-level scattering amplitudes, defined
as the on-shell limit of amputated momentum space Green's functions
of quarks, gluons and possibly other colorless particles. There has
been an impressive progress in calculating loop corrections (see e.g. \citep{Bevilacqua2013,Alioli2014})
as well as tree-level amplitudes with many external lines. 
In fact, using sophisticated computer codes, the latter can be calculated automatically
for any Standard Model process and large number of external legs \citep{Mangano:2002ea,Gleisberg:2008fv,Kleiss:2010hy,Cafarella:2007pc}.
For the purpose of the ongoing discussion, let us concentrate on the pure Yang-Mills section of the Standard Model. 
The main issue here is efficiency: due to the gluon self interactions, the number of Feynman
diagrams grows extremely fast with the order of the strong coupling
and for many external legs the expressions that have to be evaluated are enormous.
Fortunately, there are alternative methods to represent amplitudes and simplify their structure. 
The so-called color or dual decomposition \citep{Mangano:1987xk,Mangano:1990by}
is one of the essential examples. It reduces the problem to evaluating the 
color-ordered amplitudes, consisting of planar diagrams with a fixed order of the external legs. 
Further improvement of the calculation is achieved by means of recursion relations, 
for instance the so-called Berends-Giele recursion relation \citep{Berends:1987me}. 
It relates a $N$-leg off-shell current $J_{N}^{\mu}$ 
to the off-shell currents with a smaller number of legs, $J_{i}^{\mu}$, $i<N$. 
The lower-order currents are used multiple times in the recursion but need to be evaluated only once. 
The $N$-leg on-shell scattering amplitude is obtained by simply taking the on-shell limit of the current $J_{N}^{\mu}$. 

Using the Berends-Giele recursion relation together with the helicity
spinor formalism \citep{Mangano:1990by}, certain amplitudes have
been shown to have an extremely simple form for an arbitrary number of legs. 
In particular, the famous Maximally Helicity Violating (MHV) amplitudes
are expressed by just one single term \citep{Parke:1986gb}. 
The reason  that an enormous number of terms collapses into a single one  can be attributed to gauge invariance. 
Clearly, the ordinary gauge non-invariant Feynman diagrams are
not the most efficient way of calculating the gauge invariant on-shell amplitudes. 
The beautiful illustration to this issue is provided by the Britto-Cachazo-Feng-Witten (BCFW) 
recursion relation \citep{Britto:2004ap,Britto:2005fq}.
We shall recall it in some detail later on. Here, let us only remind
that the on-shell amplitude can be expressed as a sum of contributions
consisting of products of two lower order on-shell amplitudes after 
shifting two external momenta by a proper complex vector,
while preserving both on-shellness and momentum conservation.
This highly non trivial relation has its roots in the fact that gluon
amplitudes vanish when two external momenta tend to infinity in certain complex directions \citep{Arkani-Hamed2008}.
The important fact here is that the BCFW decomposition involves on-shell amplitudes, which are gauge invariant objects.
Such a decomposition proves to be very efficient.
Another more efficient way to construct amplitudes, instead of  using ordinary Feynman diagrams,
is the decomposition by means of the MHV scattering amplitudes themselves, 
which are continued analytically off-shell in a special way \cite{Cachazo2004}. 

The on-shell scattering amplitudes have some limitations though, as gluons and quarks are never free and cannot be directly 
observed in an experiment. Therefore there has been considerable interest in the systematic computation of more general objects, off-shell matrix elements, 
which of course can be reduced to the on-shell amplitudes. Such objects have a wider range of applicability, for example they can be also used in the Berends-Giele relations. 
In addition, the off-shell matrix elements are important ingredients for the so-called $k_T$ (or high energy) factorization  \cite{Catani:1990xk,Catani:1990eg,Collins:1991ty}, 
which is recently gaining more attention in the context of phenomenology of hadronic collisions. The reason is that such generalized factorization theorems, 
together with off-shell matrix elements and the so-called unintegrated parton densities, allow for a more accurate description of the kinematics of different processes, 
in particular in cases with more exclusive final states. The off-shell  scattering amplitudes, however, must be computed with a method that guarantees their gauge invariance. 
A systematic way of constructing off-shell gauge invariant amplitudes for high energy scattering was constructed in \citep{Lipatov:1995pn} and later on more efficient methods to calculate such 
amplitudes were developed  \cite{vanHameren:2012if,Kotko:2014aba,vanHameren:2014iua}. 
Let us note that the high energy off-shell amplitudes are  different from the ones used in the Berends-Giele relation. 
Namely, in the former case the off-shell gluons have longitudinal polarization, while in the Berends-Giele relation also physical transverse polarizations are involved. 
It turns out that off-shell currents with any polarization can be  extended to be gauge invariant using matrix elements
of straight Wilson lines \citep{Kotko:2014aba}, with the slope of the Wilson line corresponding to the polarization vector of the off-shell gluon. 
This construction guarantees that the Ward identities are satisfied with respect to the on-shell legs.

A parallel approach to the computation of amplitudes is realized by the light-front quantization formalism \citep{Kogut1970,Brodsky1998}. 
The corresponding light-front Feynman rules involve by construction only on-shell
lines (the off-shell propagators are traded for energy denominators).
In Refs.~\citep{Motyka2009,Cruz-Santiago2013} the light-front techniques
were used to solve the light-front equivalent of the Berends-Giele relation 
(the so-called cluster decomposition \citep{Brodsky1986})
for certain helicity configurations (see also \citep{Cruz-Santiago2015a} for a review of the light-front methods for amplitudes). 
As a result, off-shell amplitudes with one off-shell leg and in the MHV-like configuration were obtained.
Curiously, the solutions for these off-shell scattering amplitudes obtained in Refs. \citep{Cruz-Santiago2013} 
using the light-front methods feature a structure which resembles the BCFW recursion.

In the present paper we shall elaborate on the solutions and recursions obtained in Refs. \citep{Cruz-Santiago2013} and demonstrate
that they are indeed of BCFW type, but in a more general sense.
Here comes the main and novel result of the present work: 
we discover that it is a complex shift of the Wilson line slope in the matrix element that renders the
relations previously found using light-front techniques in \citep{Cruz-Santiago2013}, in a similar way as the complex shift of the particle momenta renders the BCFW relations for the on-shell case.
This method allows for further generalizations of the results of \citep{Motyka2009,Cruz-Santiago2013} to the case of different helicities.
The similarity to BCFW is structural, since the objects involved in this recursion are different:
the recursion includes both gauge dependent and gauge invariant off-shell objects, as we shall elaborate later on.
Another major difference of the derived  relations with respect to the usual BCFW recursion is the presence of additional terms,
which are due to the boundary contour integral and are not vanishing in the present case.

The issue of boundary terms in BCFW has received significant attention since the inception
of the result for pure Yang-Mills, particularly by Feng and collaborators.
For instance, \cite{Jin:2015pua} provides a nice conceptual interpretation
of boundary contributions in the BCFW recursion in terms of form factors of 
composite operators appearing in the operator product expansion of the deformed amplitude.
Also interesting recursive algorithms have been obtained for the computation of boundary terms 
for on-shell scattering amplitudes \cite{Feng:2014pia,Jin:2014qya}: the idea is to exploit the same philosophy
of the BCFW recursion relation, by studying the pole structure of boundary terms and then pinning them down
by additional complex shifts. Neither algorithm works for the most general field theory, especially
if negative coupling constants are present, as for example in effective field theories, 
but as far as Standard Model-type theories are concerned, they work fine.
Neither of these algorithms, however, can be applied to scattering amplitudes with off-shell legs. 
Concerning this issue, we must mention the results in \cite{Feng:2011twa}, where
BCFW was studied for gluon off-shell currents with and without an on-shell fermion pair, 
addressing the issue of gauge-dependence of the results and the kind of boundary terms 
which show up in case the fermion pair is deformed.

The present paper is organized as follows. First, we shall review the Berends-Giele
and BCFW recursion relations (Section~\ref{sec:RecRel}), as well
as the definition of the off-shell gauge invariant amplitudes using Wilson
lines (Section~\ref{sec:WilsonLines}) as these are the starting
points for the derivation of  our main result. In Section~\ref{sec:Derivations}
we show how to implement the complex shift of the Wilson line slope and 
we derive the relations connecting off-shell currents with fixed helicities with corresponding gauge invariant off-shell amplitudes. 
Next, in Section~\ref{sec:Applications} we apply the formulae to certain specific helicity configurations and demonstrate how
the relations previously obtained on the light-front can be recovered. 
Finally, in Section~\ref{sec:Discussion} we shall discuss more generally the meaning of the relations obtained in the present work.

\section{Recursion relations}
\label{sec:RecRel}

In order to set up the notation and set the stage for the main
results of the paper, we shall review in this Section a few key results.
 
\begin{figure}
\noindent \begin{raggedright}
\begin{minipage}[c]{0.15\paperwidth}%
\begin{center}
\includegraphics[width=0.15\paperwidth]{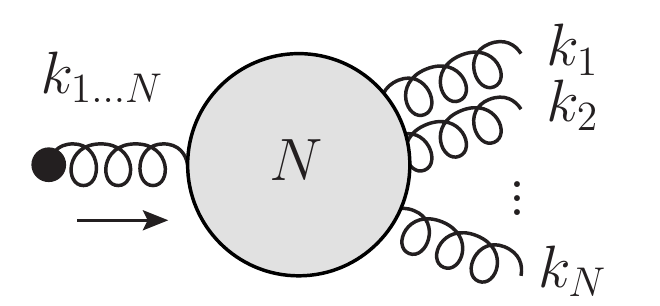}
\par\end{center}%
\end{minipage}$\,\,\,\,\,\,\,\,\,\,\,\,\,\,\,\,\,\,\,\,\,\,\,\,\,\,\,\,\,\,\,\,\,\,\,\,\,\,\,\,\,\,\,\,\,\,$%
\begin{minipage}[c]{0.45\columnwidth}%
\begin{center}
\includegraphics[width=0.45\paperwidth]{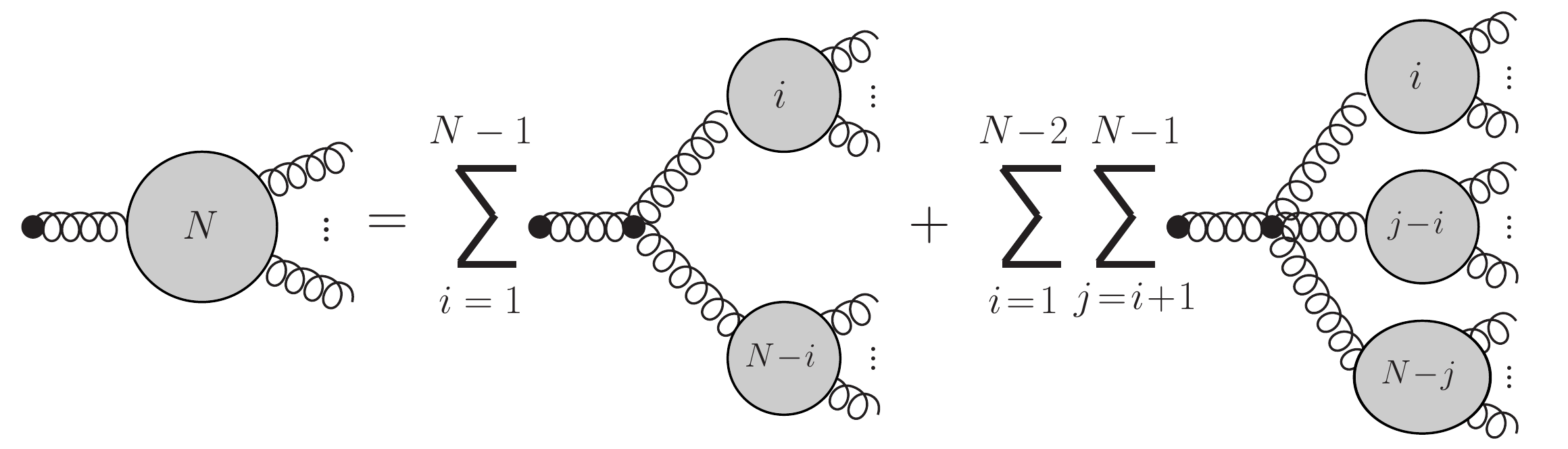}
\par\end{center}%
\end{minipage}
\par\end{raggedright}

\caption{Left: off-shell color-ordered current $J^{\mu}\left(k_{1\dots N}\right)$.
Right: the Berends-Giele recursion relation for off-shell currents.\label{fig:Berends-Giele}}

\end{figure}

Let us start by recalling the Berends-Giele recursion relation for gluons.
For a detailed review we refer to \citep{Mangano:1990by}. 
For a color-ordered off-shell current (see Fig.~\ref{fig:Berends-Giele}, left) with gluon
polarization vectors $\varepsilon_{i}^{\lambda_{i}}$ ($\lambda_{i}=\pm$)
and subject to the momentum conservation $k_{1\dots N}=k_{1}+\dots+k_{N}$,
\begin{equation}
J^{\mu}\left(k_{1\dots N};\varepsilon_{1}^{\lambda_{1}},\dots,\varepsilon_{N}^{\lambda_{N}}\right)\equiv J^{\mu\,\left(\lambda_{1}\dots\lambda_{N}\right)}\left(k_{1\dots N}\right)\,,
\end{equation}
the recursion relation reads
\begin{multline}
J^{\mu\,\left(\lambda_{1}\dots\lambda_{N}\right)}\left(k_{1\dots N}\right) = 
\frac{-i}{k_{1\dots N}^{2}}\Bigg\{
\sum_{i=1}^{N-1}V_{3}^{\mu\alpha\beta}\left(k_{1\dots i},k_{\left(i+1\right)\dots N}\right)
J_{\alpha}^{\left(\lambda_{1}\dots\lambda_{i}\right)}\left(k_{1\dots i}\right)J_{\beta}^{\left(\lambda_{i+1}\dots\lambda_{N}\right)}\left(k_{\left(i+1\right)\dots N}\right)
\\
+\sum_{i=1}^{N-2}\sum_{j=i+1}^{N-1}V_{4}^{\mu\alpha\beta\gamma}\left(k_{1\dots i},k_{\left(i+1\right)\dots j},k_{\left(j+1\right)\dots N}\right)
\\
\times J_{\alpha}^{\left(\lambda_{1}\dots\lambda_{i}\right)}\left(k_{1\dots i}\right)
J_{\beta}^{\left(\lambda_{i+1}\dots\lambda_{j}\right)}\left(k_{\left(i+1\right)\dots j}\right)J_{\gamma}^{\left(\lambda_{j+1}\dots\lambda_{N}\right)}\left(k_{\left(j+1\right)\dots N}\right)\Bigg\} \, .
\label{eq:BerendsGiele1}
\end{multline}
Here $V_{3}$ and $V_{4}$ are three-point and four-point gluon color-ordered
vertices; the independent momenta provided as arguments are outgoing, 
the remaining one incoming,
\begin{equation}
V_{3}^{\mu\alpha\beta}\left(k,p\right)=ig\left[g^{\mu\alpha}\left(-2k-p\right)^{\beta}+g^{\alpha\beta}\left(k-p\right)^{\mu}+g^{\mu\beta}\left(2p+k\right)^{\alpha}\right] \, ,
\label{eq:V3}
\end{equation}
\begin{equation}
V_{4}^{\mu\alpha\beta\gamma}\left(k,p,q\right)=ig^{2}\left(2g^{\mu\beta}g^{\alpha\gamma}-g^{\mu\alpha}g^{\beta\gamma}-g^{\mu\gamma}g^{\alpha\beta}\right)\, .
\label{eq:V4}
\end{equation}
The one-leg currents are defined as $J^{\mu}\left(k_{i};\varepsilon_{i}^{\lambda_{i}}\right)=\varepsilon_{i}^{\lambda_{i}\mu}$.
This recursion relation is schematically illustrated in Fig.~\ref{fig:Berends-Giele} (right).
The on-shell amplitude with all outgoing momenta is obtained from (\ref{eq:BerendsGiele1}) by means of the following reduction formula,
\begin{equation}
\mathcal{M}^{\left(\lambda_{0}\lambda_{1}\dots\lambda_{N}\right)}\left(k_{0},k_{1},\dots,k_{N}\right) = 
\left.ik_{1\dots N}^{2}\,\varepsilon_{0}^{\lambda_{0}}\cdot J^{\left(\lambda_{1}\dots\lambda_{N}\right)}\left(k_{1\dots N}\right)\right|_{k_{1\dots N}=-k_{0}} \, ,
\label{eq:M-onshell}
\end{equation}
where $k_{0}^{2}=0$.

In order to make  contact with the previous works \citep{Motyka2009,Cruz-Santiago2013,Cruz-Santiago2015}
which use the light-front quantization techniques, let us consider
the recursion (\ref{eq:BerendsGiele1}) in the light-cone gauge.
To this end we introduce the light cone variables by defining two null vectors
\begin{gather}
\eta=\left(1,0,0,-1\right)\,,\label{eq:eta}\\
\tilde{\eta}=\left(1,0,0,1\right)\,,\label{eq:etatild}
\end{gather}
which allow to decompose any four vector $u$ as
\begin{equation}
u^{\mu}=\frac{1}{2}u^{+}\tilde{\eta}^{\mu}+\frac{1}{2}u^{-}\eta^{\mu}+u_{\perp}^{\mu}\,,\label{eq:LCdecomp}
\end{equation}
with $u_{\perp}^{\mu}=\left(0,u^{1},u^{2},0\right)\equiv\left(0,\vec{u}_{\perp},0\right)$.
In the chosen gauge, the polarization vectors have the form
\begin{equation}
\varepsilon_{i}^{\pm\mu}=\varepsilon_{\perp}^{\pm\mu}+\frac{\vec{\varepsilon}_{\perp}^{\,\,\pm}\cdot\vec{k}_{i\perp}}{k_{i}\cdot\eta}\eta^{\mu}\,,\label{eq:LFpolvec1}
\end{equation}
where 
\begin{equation}
\varepsilon_{\perp}^{\pm\mu}=\frac{1}{\sqrt{2}}\left(0,1,\pm i,0\right)\,.\label{eq:LFpolvec2}
\end{equation}
They satisfy the conditions $k_{i}\cdot\varepsilon_{i}^{\pm}=\eta\cdot\varepsilon_{i}^{\pm}=\varepsilon_{i}^{\pm}\cdot\varepsilon_{i}^{\pm}=0$
and $\varepsilon_{i}^{\pm}\cdot\varepsilon_{i}^{\mp}=-1$. 
Let us note that these relations hold irrespectively whether the momentum $k_{i}$ is on-shell
or off-shell. In the gauge we are using, the numerator of a propagator is 
\begin{equation}
\sum_{\lambda=\pm}\varepsilon_{i\mu}^{\lambda}\varepsilon_{i\nu}^{\lambda*}=d_{\mu\nu}\left(k_{i},\eta\right)=-g_{\mu\nu}+\frac{k_{i\,\mu}\eta_{\nu}+\eta_{\mu}k_{i\,\nu}}{k_{i}\cdot\eta} 
- \frac{\eta_{\mu}\eta_{\nu}}{(k_i\cdot\eta)^2}k_i^2 \,. \label{LCgauge}
\end{equation}

We may write a light-cone gauge analog of (\ref{eq:BerendsGiele1})
as follows (for similar approach see \citep{Kosower1990}). For any off-shell current $J^{\mu}$ we change the propagators to the light-cone gauge propagators (\ref{LCgauge}). 
This obviously changes the currents, as they are not gauge invariant, but does not change the form of (\ref{eq:BerendsGiele1}). 
In what follows, we will refer to these new currents with the same symbol, as we will be dealing with them all through the rest of the paper.
For the new light-cone gauge currents we may write 

\begin{equation}
J^{\left(\lambda_{1}\dots\lambda_{i}\right)\, \mu }\left(k_{1\dots i}\right) = \sum_{\lambda_{1\dots i} = 
\pm}\varepsilon_{1\dots i}^{\lambda_{1\dots i}\, \mu}\, J^{\left(\lambda_{1\dots i}\rightarrow\lambda_{1}\dots\lambda_{i}\right)}\left(k_{1\dots i}\right)\,.
\end{equation}
Above, the off-shell current with fixed external helicity projections is defined as 
\begin{equation}
J^{\left(\lambda_{1\dots i}\rightarrow\lambda_{1}\dots\lambda_{i}\right)}\left(k_{1\dots i}\right) = \varepsilon_{1\dots i}^{\lambda_{1\dots i}\, * \, \nu} 
\, \frac{-i}{k^2_{1\dots i} } J^{\left(\lambda_{1}\dots\lambda_{i}\right)}_{(\mathrm{amp})\,\nu}\left(k_{1\dots i}\right)\,,
\end{equation}
where $J_{(\mathrm{amp})}$ is the amputated current in the light-cone gauge. 
When the off-shell propagator $k_{1\dots i}^2$ is amputated in $J^{\left(\lambda_{1\dots i}\rightarrow\lambda_{1}\dots\lambda_{i}\right)}$,
the resulting object is called an {\it off-shell amplitude}.

Contracting (\ref{eq:BerendsGiele1}) with the polarization vector $\varepsilon^{*}_{1\dots N}$, we finally get the following 
light-cone analog of the Berends-Giele recursion for fixed helicities: 
\begin{multline}
J^{\,\left(\lambda_{1\dots N}\rightarrow\lambda_{1}\dots\lambda_{N}\right)}\left(k_{1\dots N}\right)=\frac{-i}{k_{1\dots N}^{2}}\\
\Bigg\{\sum_{\lambda,\lambda'=\pm}\,
\sum_{i=1}^{N-1}V_{3}^{\left(\lambda_{1\dots N}\rightarrow\lambda\lambda'\right)}\left(k_{1\dots i},k_{\left(i+1\right)\dots N}\right)
J^{\,\left(\lambda\rightarrow\lambda_{1}\dots\lambda_{i}\right)}\left(k_{1\dots i}\right)J^{\,\left(\lambda'\rightarrow\lambda_{i+1}\dots\lambda_{N}\right)}\left(k_{\left(i+1\right)\dots N}\right)\\
+\sum_{\lambda,\lambda',\lambda''=\pm}\,\sum_{i=1}^{N-2}
\sum_{j=i+1}^{N-1}V_{4(\mathrm{LC})}^{\left(\lambda_{1\dots N}\rightarrow\lambda\lambda'\lambda''\right)}\left(k_{1\dots i},k_{\left(i+1\right)\dots j},k_{\left(j+1\right)\dots N}\right)\\
\times J^{\,\left(\lambda\rightarrow\lambda_{1}\dots\lambda_{i}\right)}\left(k_{1\dots i}\right)J^{\,\left(\lambda'\rightarrow\lambda_{i+1}\dots\lambda_{j}\right)}
\left(k_{\left(i+1\right)\dots j}\right)J^{\,\left(\lambda''\rightarrow\lambda_{j+1}\dots\lambda_{N}\right)}\left(k_{\left(j+1\right)\dots N}\right)\Bigg\} \, . 
\label{eq:BerendsGiele2}
\end{multline}
The polarization-fixed gluon vertices in (\ref{eq:BerendsGiele2}) are defined  by contracting the three-gluon 
(\ref{eq:V3}) and  four-gluon (\ref{eq:V4}) vertices with the polarization vectors, 
the incoming leg being contracted with a complex conjugated one.
In addition, the four-gluon vertex has to be modified due to the third term of (\ref{LCgauge}).
This term is proportional to $k^2$, thus it effectively gives the instantaneous interaction terms known from the
light-front quantization. These terms can be conveniently combined with four-gluon vertices. 
Such vertices were also used  in \citep{Kosower1990}, although with a different momentum flow.
We shall give the precise form of $V_{4(\mathrm{LC})}$ in Section~\ref{sec:Applications}, 
where we use it in an actual computation.
As we mentioned, the off-shell currents defined above are not gauge invariant. 
This manifests itself as the lack of the Ward identities for on-shell legs, 
\begin{equation}
\left.J^{\,\left(\lambda\rightarrow\lambda_{1}\dots\lambda_{i}\right)}\left(k_{1\dots i}\right)\right|_{\varepsilon_{j}^{\lambda_{j}}\rightarrow k_{j}}\neq0,\,\,\, j=1,\dots,i \, .
\end{equation}
Obviously these Ward identities are restored in the on-shell limit (\ref{eq:M-onshell}).

Let us now consider two particular helicity configurations for which there  exist a few interesting results  in the literature. 
For example, consider the $+\rightarrow+\dots+$ off-shell current. 
While  the on-shell scattering amplitudes with the $\pm\rightarrow+\dots+$
helicity configurations vanish \citep{Mangano:1990by},
in the off-shell case only $-\rightarrow+\dots+$ is zero, 
whereas the other, $+\rightarrow+\dots+$, does not vanish. 
For this helicity configuration the recursion (\ref{eq:BerendsGiele2}) reduces to
\begin{multline}
J^{\,\left(+\rightarrow+\dots+\right)}\left(k_{1\dots N}\right)=\frac{-i}{k_{1\dots N}^{2}}\,\sum_{i=1}^{N-1}V_{3}^{\left(+\rightarrow++\right)}\left(k_{1\dots i},k_{\left(i+1\right)\dots N}\right)\\
\times J^{\,\left(+\rightarrow+\dots+\right)}\left(k_{1\dots i}\right)J^{\,\left(+\rightarrow+\dots+\right)}\left(k_{\left(i+1\right)\dots N}\right)\,.\label{eq:BerendsGiele3}
\end{multline}
It follows from $V_{4}^{\left(+\rightarrow+++\right)}=0$ (see e.g. Table 2 of \citep{Cruz-Santiago2015a}) and $J^{\left(-\rightarrow+\dots+\right)}=0$, as mentioned above. 
This recursion was solved in \citep{Motyka2009} using the light-front quantization approach. 
Earlier the solution had been found in \citep{Berends:1987me} and in \citep{Bardeen1996,Rosly1997} 
as a solution to the self-dual Yang-Mills theory.
The solution turns out to be very compact and reads
\begin{equation}
J^{\,\left(+\rightarrow+\dots+\right)}\left(k_{1\dots N}\right) 
= - \left(-g\right)^{N-1}\,\frac{\tilde{v}_{\left(1\dots N\right)1}}{\tilde{v}_{1\left(1\dots N\right)}}\,\frac{1}{\tilde{v}_{N\left(N-1\right)}\dots\tilde{v}_{32}\,\tilde{v}_{21}}\,,\label{eq:Mplusplus}
\end{equation}
where we have defined the following convenient notation
\begin{equation}
\tilde{v}_{\left(p\right)\left(r\right)}=p\cdot\varepsilon_{r}^{-}\label{eq:vtild} \; ,
\end{equation}
for any two momenta $p$, $r$ (for example $\tilde{v}_{ij}=k_{i}\cdot\varepsilon_{j}^{-}$,
$\tilde{v}_{\left(1\dots i\right)\left(1\dots j\right)}=k_{1\dots i}\cdot\varepsilon_{1\dots j}^{-}$ etc.).

The next less trivial example we want to discuss is the MHV  helicity configuration
$+\rightarrow-+\dots+$. For that case, the recursion (\ref{eq:BerendsGiele2}) reduces to 
\begin{multline}
J^{\,\left(+\rightarrow-+\dots+\right)}\left(k_{1\dots N}\right)=\frac{-i}{k_{1\dots N}^{2}}\\
\Bigg\{\sum_{i=1}^{N-1}V_{3}^{\left(+\rightarrow++\right)}\left(k_{1\dots i},k_{\left(i+1\right)\dots N}\right)J^{\,\left(+\rightarrow-+\dots+\right)}
\left(k_{1\dots i}\right)J^{\,\left(+\rightarrow+\dots+\right)}\left(k_{\left(i+1\right)\dots N}\right)\\
+\sum_{i=1}^{N-1}V_{3}^{\left(+\rightarrow-+\right)}\left(k_{1\dots i},k_{\left(i+1\right)\dots N}\right)
J^{\,\left(-\rightarrow-+\dots+\right)}\left(k_{1\dots i}\right)J^{\,\left(+\rightarrow+\dots+\right)}\left(k_{\left(i+1\right)\dots N}\right)\\
+\,\sum_{i=1}^{N-1}\sum_{j=i+1}^{N-2}V_{4}^{\left(+\rightarrow-++\right)}\left(k_{1\dots i},k_{\left(i+1\right)\dots j},k_{\left(j+1\right)\dots N}\right)\\
\times J^{\,\left(-\rightarrow-+\dots+\right)}\left(k_{1\dots i}\right)J^{\,\left(+\rightarrow+\dots+\right)}\left(k_{\left(i+1\right)\dots j}\right)
J^{\,\left(+\rightarrow+\dots+\right)}\left(k_{\left(j+1\right)\dots N}\right)\Bigg\}\,.\label{eq:BerendsGiele4}
\end{multline}
Unlike for the helicity configurations considered before, this recursion is much more complicated. 
Remarkably, the solution has been found using the light-front quantization techniques \citep{Cruz-Santiago2013}
and  is expressed by the following relation
\begin{multline}
J^{\,\left(+\rightarrow-+\dots+\right)}\left(k_{1\dots N}\right)=\tilde{J}^{\,\left(+\rightarrow -+\dots+\right)}\left(k_{1\dots N}\right)\\
-i g\sum_{i=2}^{N-1}\tilde{J}^{\,\left(+\rightarrow-+\dots+\right)}\left(k_{1\dots i}\right)\frac{k_{1\dots N}^{+}}{k_{\left(i+1\right)\dots N}^{+}\,
\tilde{v}_{\left(1\dots i\right)\left(i+1\right)}}\,J^{\,\left(+\rightarrow+\dots+\right)}\left(k_{\left(i+1\right)\dots N}\right)\,.\label{eq:RecRel1}
\end{multline}
The new object, $\tilde{J}$, emerges in that expression due to successive resummations of  the light-front diagrams
\footnote{
The relation in \citep{Cruz-Santiago2013} was actually obtained for  slightly different objects, namely for 
amputated off-shell currents (modulo a constant phase factor).}.
It turns out that it has a simple one-term form
\begin{equation}
\tilde{J}^{\,\left(+\rightarrow-+\dots+\right)}\left(k_{1\dots i}\right) = 
\frac{2 g^{i-1}}{k_{1\dots i}^{2}}\,\left(\frac{k_{1\dots i}^{+}}{k_{1}^{+}}\right)^{2} \,
\frac{\tilde{v}_{1\left(1\dots i\right)}^{3}}{\tilde{v}_{(1\dots i) i}\tilde{v}_{i\left(i-1\right)}\dots\tilde{v}_{32}\,\tilde{v}_{21}}\,.\label{eq:MHV}
\end{equation}
Interestingly, the structure of (\ref{eq:MHV}) is the same (modulo the propagator) as the on-shell MHV amplitude written in terms of
products of helicity spinors \citep{Mangano:1990by}. 
This is apparent in the light-front quantization formalism, where the variables $\tilde{v}_{ij}$ are expressed by products of massless helicity spinors
\footnote{Actually this connection holds here as well, but for $\tilde{v}_{ij}$
involving on-shell particles only. In the light-front quantization
all lines are on-shell and this identification holds for any
line.%
} \citep{Motyka2009,Cruz-Santiago2013}. 
One has to keep in mind that (\ref{eq:MHV}) is an off-shell object, i.e. the momentum
$k_{1\dots i}^{2} \neq 0 $. This remarkable fact was a trigger for further
studies of the gauge invariance properties of the relation (\ref{eq:RecRel1}).
It was proved in \citep{Cruz-Santiago2015} that the reason why the auxiliary
object (\ref{eq:MHV}) appearing in the solution to (\ref{eq:BerendsGiele4})
has this extremely simple form is that it is gauge invariant. 
More precisely, (\ref{eq:MHV}) turns out to be directly related to the \textit{gauge invariant
off-shell amplitude}.
These objects are commonly used in high energy scattering processes \citep{vanHameren:2012if,vanHameren:2014iua,vanHameren:2012uj,vanHameren:2015bba}, 
but the idea can also be extended beyond the high-energy limit of QCD. 
In Ref.~\citep{Kotko:2014aba} a construction of tree-level gauge invariant amplitudes with one or more off-shell legs was formulated
in terms of matrix elements of straight infinite Wilson line operators.
In \citep{Cruz-Santiago2015} it was shown that (\ref{eq:MHV}) originates
in the gluonic matrix elements of such Wilson line operators and thus is gauge invariant 
-- which was also demonstrated explicitly by verifying  the Ward identities
\begin{equation}
\left.\tilde{J}^{\,\left(\lambda_{1\dots i}\rightarrow\lambda_{1}\dots\lambda_{i}\right)}\left(k_{1\dots i}\right)\right|_{\varepsilon_{j}^{\lambda_{j}}\rightarrow k_{j}}=0,\,\,\, j=1,\dots,i \, .
\end{equation}

The relation (\ref{eq:RecRel1}) has also one more intriguing property.
Namely, when the sum on the r.h.s is moved to the l.h.s,
the resulting recursion for $\tilde{J}$ has a similar structure
to the BCFW recursion. 
However, with respect to the latter it is modified by the presence of another $N$-leg term, the current $J$. 
Actually, such terms are expected to show up in the BCFW construction as boundary contributions, 
in the event that the amplitude does not vanish for the complex auxiliary parameter $z$ being taken to infinity.

To better understand the above statements, let us recall the derivation of the  BCFW
recursion in some detail \citep{Britto:2004ap,Britto:2005fq}. In order to obtain the relation for an
on-shell amplitude $\mathcal{M}^{\left(\lambda_{1}\dots\lambda_{N}\right)}\left(k_{1},\dots,k_{N}\right)$,
one shifts two external on-shell momenta $i,j$ by a complex four
vector $z\, e^{\mu}$, with $e^{\mu}$ being a fixed four vector and $z$ a complex number, 
such that they remain on-shell and that total momentum conservation is preserved. 
Let us consider a specific case where we  shift the momenta $1$ and $N$. 
We can construct a meromorphic function
\begin{equation}
f\left(z\right)=\frac{1}{z}\,\mathcal{M}^{\left(\lambda_{1}\dots\lambda_{N}\right)}\left(k_{1}+ze,\dots,\dots,k_{N}-ze\right)\,,
\end{equation}
which for $\left|z\right|\rightarrow\infty$ 
vanishes like $f\left(z\right)\sim\mathcal{O}\left(\left|z\right|^{-2}\right)$ (or even faster) for a proper choices of the shift vector $e^\mu$.

This fact is by no means obvious and was first proved in \citep{Britto:2005fq} for pure Yang-Mills without and with fermions
and more deeply explored in \citep{Arkani-Hamed2008} also for more general theories. 
Thanks to it, the integral of $f\left(z\right)$ over a circle at infinity vanishes and the residue at $z=0$, 
which corresponds to the original (non-shifted) amplitude, can be expressed
as the sum over the remaining residues, which are relatively simple to calculate. 
The residues are calculated at the poles $z_{m}$ originating
in propagators which diverge when the shifted momentum joining two sub-diagrams becomes on-shell. 
The scattering amplitude in the BCFW decomposition can be then written as a sum over the residues as follows
\begin{multline}
\mathcal{M}^{\left(\lambda_{1}\dots\lambda_{N}\right)}\left(k_{1},\dots,k_{N}\right)=\sum_{m=2}^{N-3}\sum_{\lambda_{m}=
\pm}\mathcal{M}^{\left(\lambda_{1}\dots\lambda_{m}\right)}\left(k_{1}+z_{m}e,\dots,k_{1\dots m-1}+z_{m}e\right)\\
\times\frac{1}{k_{1\dots m-1}^{2}}\,\mathcal{M}^{\left(-\lambda_{m}\dots\lambda_{N}\right)}\left(-k_{1\dots m-1}-z_{m}e,\dots,k_{N}-z_{m}e\right)\,,\label{eq:BCFW}
\end{multline}
where the poles are equal to $z_{m}=k_{1\dots m-1}^{2}/\left(2e\cdot k_{1\dots m-1}\right)$.
Here a few comments are in order. The amplitudes on the r.h.s. of this
relation involve complex momenta but are on-shell and thus are gauge invariant.
It is the most important difference between (\ref{eq:BerendsGiele1})
and (\ref{eq:BCFW}). The former does not involve gauge invariant
quantities and, thus, typically has many more terms then appearing in (\ref{eq:BCFW}). 
Let us stress that the relation (\ref{eq:BCFW}) emerges because the amplitude with shifted momenta vanishes at infinity; 
I either a surface term would be present or the integral over $dz$ over the circle at infinity would be divergent.
Such a boundary term would have $N$ legs, and -- since the residue at $z=0$ has $N$ legs as well -- the recursion would not emerge.
This looks like what happened in (\ref{eq:RecRel1}). 
However, QCD amplitudes vanish for infinite momenta at some particular
complex directions and thus there are no boundary terms. 
This fact remains true for gauge invariant off-shell amplitudes as well \citep{vanHameren:2014iua,vanHameren:2015bba}.

Thus, the question we ask and answer in the present paper is: is it possible to recover (\ref{eq:RecRel1}) by applying  
Cauchy's theorem to a quantity depending on some variables which get shifted into the complex plane?
It turns out that it is possible and the quantity undergoing the complex shift 
is the direction of the Wilson line used to define gauge invariant off-shell amplitudes. 
We will elaborate on this in the following sections.

\section{Gauge invariant off-shell currents}
\label{sec:WilsonLines}

As we mentioned, the helicity-fixed off-shell currents appearing in the recursion
relation (\ref{eq:BerendsGiele2}) are not gauge invariant. 
One can however find a proper gauge invariant extension using matrix elements of Wilson line operators \citep{Kotko:2014aba}. 
We have already encountered such object in Eq.~(\ref{eq:RecRel1}). 
Below we shall briefly recall the basic ideas more systematically.

Let us consider the following matrix element
\begin{equation}
\mathfrak{M}=\int\!\! d^{4}x\, e^{ik\cdot x}\left\langle 0\left|\mathcal{T}
\left\{ R_{\mathfrak{e}}^{\, a}(x)\, e^{iS_{\,\textrm{Y-M}}}\right\} \right|k_{1},\lambda_{1},a_{1};\ldots;k_{N},\lambda_{N},a_{N}\right\rangle _{\mathrm{c}} \, ,
\label{eq:Mfrak}
\end{equation}
where 
\begin{equation}
R_{\mathfrak{e}}^{\, a}(x)=\frac{1}{\pi g} \,\mathrm{Tr}\left[t^{a}\mathcal{P}\exp\left(ig\int_{-\infty}^{+\infty}ds\, A_{\mu}^{b}\left(x+s\,\mathfrak{e}\right)\mathfrak{e}^{\mu}t^{b}\right)\right] \, ,
\label{eq:Roperator}
\end{equation}
is the Wilson line operator. In the above definition, $\mathcal{T}$ is the time-ordering,
$\mathcal{P}$ is the path-ordering, $S_{\,\textrm{Y-M}}$ is the
Yang-Mills  action, and $\left|k_{i},\lambda_{i},a_{i}\right\rangle $
are one-gluon on-shell states with momentum $k_{i}$, helicity $\lambda_{i}$
and color $a_{i}$. The subscript `$\mathrm{c}$' means that we take
only connected contributions. The path ordered exponential in the
Wilson line operator  is defined to be an infinite
straight line parametrized as $z^{\mu}\left(s\right)=x^{\mu}+s\mathfrak{e}^{\mu}$ with fixed four-vector $\mathfrak{e}$.
Such operator is explicitly gauge invariant with respect to  small
gauge transformations and consequently such is its matrix element involving
on-shell external particles. The interpretation of the operator $R_{\mathfrak{e}}^{\, a}$
is such that its Fourier transform creates a `physical' dressed off-shell
gluon state with color charge $a$, `polarization' $\mathfrak{e}$
and momentum $k$, $k^{2}\neq0$. Consequently (\ref{eq:Mfrak}) is
related to the scattering amplitude of such a state to produce $N$
on-shell gluon states, more precisely 
\begin{equation}
\mathfrak{M}=\delta^{4}\left(k-k_{1}-\ldots-k_{N}\right)\delta\left(\mathfrak{e}\cdot k\right)\tilde{J}^{\,\left(\mathfrak{e}\rightarrow\lambda_{1}\dots\lambda_{N}\right)}\left(k\right)\,,\label{eq:Mfrak_2}
\end{equation}
where $\tilde{J}$ 
is the {\it gauge invariant off-shell current} \footnote{\label{footnote3} We change the notation here comparing to Refs.~\citep{Kotko:2014aba,Cruz-Santiago2015} in order to avoid further proliferation of 
notation when introducing new objects needed in the present work. In fact, in the present paper we use the more precise term: `gauge invariant off-shell current' instead 
of `gauge invariant off-shell amplitude'. 
The relation of the new notation to the one in \citep{Cruz-Santiago2015} is $\tilde{J}=(ig)^{-1}\tilde{\mathcal{M}}$.} for a transition of a `dressed' gluon with `polarization' vector $\mathfrak{e}$ to $N$ on-shell gluon states with polarization vectors $\varepsilon^\lambda_1,\dots ,\varepsilon^\lambda_N$.
It is the same object appearing in the recursion (\ref{eq:RecRel1}) as demonstrated in \citep{Cruz-Santiago2015}.  
It can be shown that the Ward identity for the 
current
$\tilde{J}$ is proportional to the product $k\cdot\mathfrak{e}$,
\begin{equation}
\left.\tilde{J}^{\, \left(\mathfrak{e}\rightarrow \lambda_{1}\dots\lambda_{N}\right)}\left(k\right)\right|_{\varepsilon_{i}\rightarrow k_{i}}\sim k\cdot\mathfrak{e},\label{eq:Mtild_WI}
\end{equation}
and thus vanishes if the `polarization' vector $\mathfrak{e}$ is
transverse to $k$. This is assured for the whole matrix element $\mathfrak{M}$
by the Dirac delta $\delta\left(\mathfrak{e}\cdot k\right)$. 
Let us note that the objects in (\ref{eq:Mfrak_2}) actually carry color indices.
They were omitted for brevity as in the present work we always work with color stripped quantities.
We also set $k=k_{1\dots N}=k_{1}+\dots+k_{N}$.

Diagrammatically, the color-ordered off-shell current $\tilde{J}$
can be expressed as

\begin{flushleft}
\begin{tabular}{>{\centering}m{0.87\columnwidth}>{\centering}m{0.05\columnwidth}}
\bigskip{}

\raggedright{}\includegraphics[height=0.06\paperheight]{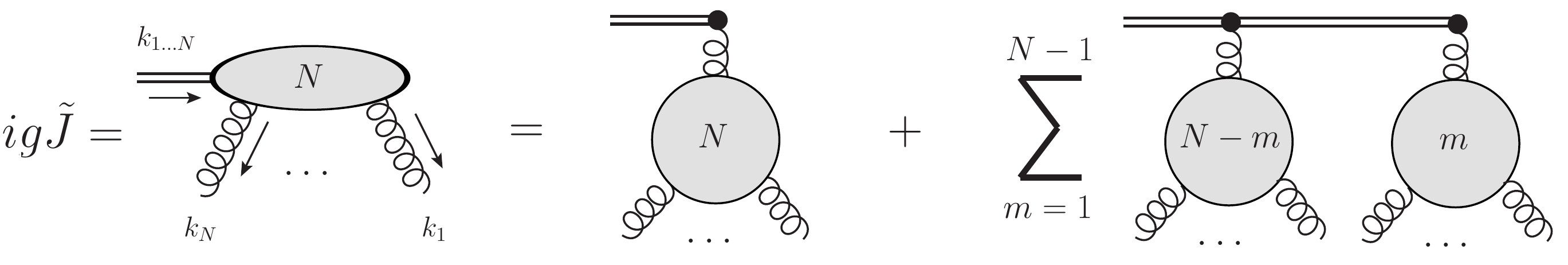} & \tabularnewline
\end{tabular}
\par\end{flushleft}

\begin{flushleft}
\begin{tabular}{>{\centering}m{0.87\columnwidth}>{\centering}m{0.05\columnwidth}}
\raggedleft{}\includegraphics[height=0.06\paperheight]{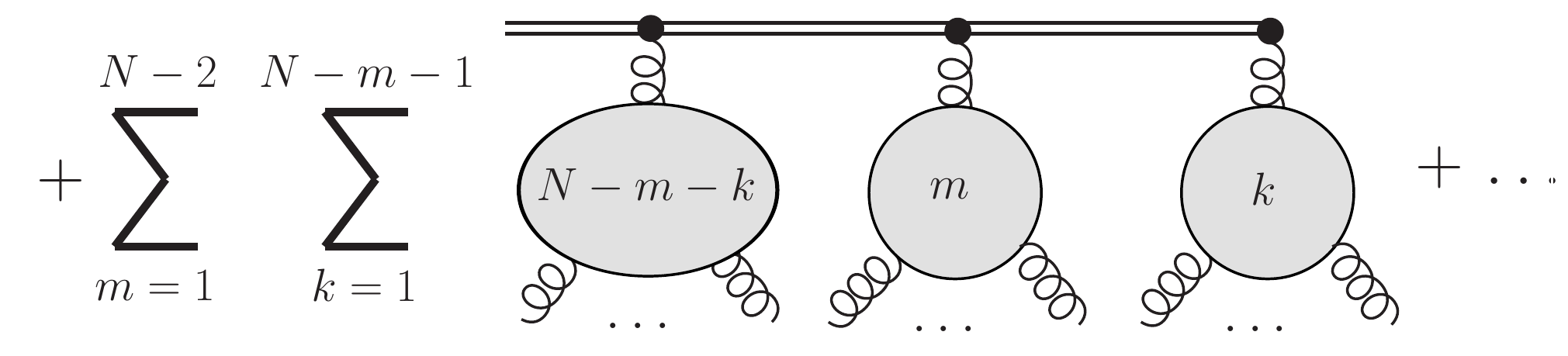} & \centering{}\centering{(\myref )}
\refmyref{Gaugelink_expansion}\tabularnewline
\end{tabular}
\par\end{flushleft}

\noindent \noindent The double line represents a propagation of the
dressed off-shell gluon state (the Wilson line). Its propagator has
the eikonal form $i/\left(p\!\cdot\!\mathfrak{e}+i\epsilon\right)$,
with $p$ being the momentum flowing through the line. The gluons
couple to the Wilson line (the double-line) via an $ig\mathfrak{e}^{\mu}$
vertex. The blobs represent the standard QCD contributions with the
numbers indicating how many  external on-shell legs each of them features;
indeed, they are precisely the off-shell currents $J\cdot\mathfrak{e}$ of Fig.~\ref{fig:Berends-Giele}.
The ellipses after the last plus sign represent contributions with
more blobs connected to the double line. More details on the above construction can be found in \citep{Kotko:2014aba}.

The gauge invariance condition for $\tilde{J}$,  $k_{1\dots N}\cdot\mathfrak{e}=0$,
is satisfied by the polarization vector $\varepsilon_{1\dots N}^{\lambda_{1\dots N}}$ defined in (\ref{eq:LFpolvec1}). 
Thus all the $\tilde{J}^{\,\left(+\rightarrow -+\dots+\right)}\left(k_{1\dots i}\right)$ currents in (\ref{eq:RecRel1}) are gauge invariant. 
Let us note that, while the whole matrix element $\mathfrak{M}$ is rather ill-defined for $\mathfrak{e} = \varepsilon_{1\dots N}^{\lambda_{1\dots N}}$ 
($\mathfrak{e}$ should be a constant and momentum independent vector, 
so that  $\mathfrak{M}$ is a generalized function in the component of the off-shell momentum along   $\mathfrak{e}$), 
the quantity of interest, i.e. $\tilde{J}$, is a perfectly valid object.

We will often encounter yet another class of objects, superficially similar to the above gauge invariant currents $\tilde{J}$.
Imagine the diagrams (\arabic{Gaugelink_expansion}) with some $\mathfrak{e}$
which \textit{does not} satisfy $\mathfrak{e}\cdot k_{1\dots N}=0$;
this is in principle possible, as the definition of $\tilde{J}$ via the diagrams (\arabic{Gaugelink_expansion}) does not require that $\mathfrak{e}$ is transverse.
Then, obviously, the Ward identity (\ref{eq:Mtild_WI}) will not be fulfilled and such $\tilde{J}$  will not be gauge invariant.
For example, we shall encounter objects where $\mathfrak{e}=\varepsilon_{1\dots m}$, $m>i$. 
Then $k_{1\dots i}\cdot \varepsilon_{1\dots m} \neq 0$ and $\tilde{J}^{\,\left(\varepsilon_{1\dots m}\rightarrow -+\dots+\right)}\left(k_{1\dots i}\right)$ is not gauge invariant.
For a latter purpose, we find it is convenient to re-define such non-gauge-invariant currents to include the leftmost double-line propagator:
\begin{equation}
\tilde{J}'^{\, \left(\mathfrak{e}\rightarrow \lambda_{1}\dots\lambda_{N}\right)}
\left(k_{1\dots N}\right)=\frac{i}{-k_{1\dots N}\cdot\mathfrak{e}}\,\tilde{J}^{\, \left(\mathfrak{e}\rightarrow \lambda_{1}\dots\lambda_{N}\right)}
\left(k_{1\dots N}\right).\label{eq:Mtildbar}
\end{equation}
We note that it is possible to define the above quantity in terms of matrix elements of a Wilson line operator, similar to (\ref{eq:Mfrak}), (\ref{eq:Roperator}),
but with the semi-infinite path, e.g. spanning from $0$ to $\infty$.
However, since in the present work we actually never use the direct definitions but rather the diagrammatic expressions
like (\arabic{Gaugelink_expansion}), we will not go into a precise definition here. 

\section{Derivations}
\label{sec:Derivations}

\subsection{Prerequisities and the basic idea}
\label{sub:BasicIdea}

In the following subsections we will derive the BCFW-like recursion
relations we mentioned in Section~\ref{sec:RecRel}.
The basic idea is to investigate the analytic structure of $\tilde{J}$
(or $\tilde{J}'$) when we perform a complex shift of the Wilson line slope 
\begin{equation}
\mathfrak{e}^{\mu}\rightarrow\mathfrak{e}^{\mu}\left(z\right)=\mathfrak{e}^{\mu}+z\mathfrak{e}'^{\mu}\,,\label{eq:shift1}
\end{equation}
with a certain fixed four vector $\mathfrak{e}'$. 
To this end, let us put the diagrammatic expression (\arabic{Gaugelink_expansion})
into an algebraic form,
\begin{eqnarray}
i\, g \tilde{J}^{\, \left(\mathfrak{e}(z)\rightarrow \lambda_{1}\dots\lambda_{N}\right)} \left(k_{1\dots N}\right)
&=& 
i g\,\mathfrak{e}\left(z\right) \cdot J^{\left(\lambda_{1}\dots\lambda_{N}\right)}\left(k_{1\dots N}\right)
\nonumber \\
&& \hspace{-25mm}
+ \sum_{i=1}^{N-1}\left(ig\right)^{2}\mathfrak{e}\left(z\right)\cdot J^{\left(\lambda_{i+1}\dots\lambda_{N}\right)}
\left(k_{\left(i+1\right)\dots N}\right)\,\frac{i}{-k_{1\dots i}\cdot\mathfrak{e}\left(z\right)}\,\mathfrak{e}\left(z\right)\cdot J^{\left(\lambda_{1}\dots\lambda_{i}\right)}\left(k_{1\dots i}\right)
\nonumber \\
&& \hspace{-25mm}
+ \sum_{i=1}^{N-2}\sum_{j=i+1}^{N-1}\left(ig\right)^{3}\mathfrak{e}\left(z\right)\cdot J^{\left(\lambda_{j+1}\dots\lambda_{N}\right)}
\left(k_{\left(j+1\right)\dots N}\right)\,\frac{i}{-k_{1\dots j}\cdot\mathfrak{e}\left(z\right)}\,\mathfrak{e}\left(z\right)\cdot J^{\left(\lambda_{i+1}\dots\lambda_{j}\right)}\left(k_{\left(i+1\right)\dots j}\right)
\nonumber \\
&& \hspace{-25mm}
\times \frac{i}{-k_{1\dots i}\cdot\mathfrak{e}\left(z\right)}\,\mathfrak{e}\left(z\right)\cdot J^{\left(\lambda_{1}\dots\lambda_{i}\right)}\left(k_{1\dots i}\right)+\dots\label{eq:Mtildz1}
\end{eqnarray}
We see that, superficially, $\tilde{J}^{\, \left(\mathfrak{e}(z)\rightarrow \lambda_{1}\dots\lambda_{N}\right)}$
is linear in $\mathfrak{e}\left(z\right)$ and thus behaves like $\mathcal{O}\left(z^{1}\right)$, 
whereas $\tilde{J}'$ would behave like $\mathcal{O}\left(z^{0}\right)$. 
However, this critically depends on the choice of the vector $\mathfrak{e}'$, i.e. the direction of the shift. 
For example, let us consider a choice for which we could have
\begin{equation}
\mathfrak{e}'\cdot J^{\left(\lambda_{i}\dots\lambda_{i+p}\right)}\left(k_{i\dots\left(i+p\right)}\right)=0,\,\,\,\, \;\;i=1,\dots,N,\,\,\,\,\; \;p=0,\dots,N-i.\label{eq:Jecond}
\end{equation}
Is that choice possible for a constant, momentum independent $\mathfrak{e}'$? 
The answer is positive, but  the choice is not universal. For example, if all the helicity projections are the same, 
e.g. $\lambda_{1}=\dots=\lambda_{N} = +$ and either the polarization vectors are chosen as in (\ref{eq:LFpolvec1}) 
or they have the same reference momentum, we may choose $\mathfrak{e}'=\varepsilon_{p}^{+}$ where $p$ is any momentum. This works in any gauge. 
A better choice, suitable for any helicity configuration but working only in the light-cone gauge introduced in Section~\ref{sec:RecRel} is 
\begin{equation}
\mathfrak{e'^{\mu}=\eta^{\mu}} \, ,\label{eq:echoice}
\end{equation}
which satisfies (\ref{eq:Jecond}) because $d_{\mu\nu}\left(p,\eta\right)\eta^{\nu}=0$
(recall that the off-shell currents $J$ include propagators and thus the $d_{\mu\nu}$ projectors). 
For the choice (\ref{eq:echoice}) and in the light-cone gauge, one finds
$$
\tilde{J}^{\, \left(\mathfrak{e}(z)\rightarrow \lambda_{1}\dots\lambda_{N}\right)} \sim \mathcal{O}\left(z^{0}\right)
\quad \text{and} \quad  
\tilde{J}'^{\, \left(\mathfrak{e}(z)\rightarrow \lambda_{1}\dots\lambda_{N}\right)} \sim \mathcal{O}\left(z^{-1}\right) \, .
$$
 We underline that they are gauge-dependent quantities, as discussed at the end of Section~\ref{sec:WilsonLines}.

\subsection{The general relations}
\label{sub:Relations}

Let us define a meromorphic function
\begin{equation}
f\left(z\right)=\frac{1}{z}\,\tilde{J}^{\, \left(\mathfrak{e}(z)\rightarrow \lambda_{1}\dots\lambda_{N}\right)} \, ,
\end{equation}
with the choice of $\mathfrak{e}\left(z\right)$ given in (\ref{eq:shift1}), (\ref{eq:echoice}), i.e.
\begin{equation}
\mathfrak{e}^{\mu}\left(z\right)=\mathfrak{e}^{\mu}+z\eta^{\mu}\,,\label{eq:shift111}
\end{equation}
and $\tilde{J}$ defined in the light-cone gauge. As discussed
in the previous section, with that choice the only $z$ dependence
is in the denominators, more precisely in the double-line propagators of (\arabic{Gaugelink_expansion}). 
Thus the analytic structure of $f\left(z\right)$ is extremely simple: besides the obvious
pole at $z=0$ we have $N-1$ poles, $z_{1\dots i}$, corresponding
to vanishing  of a  Wilson line (the double-line) propagators
\begin{equation}
k_{1\dots i}\cdot\mathfrak{e}\left(z\right)=0\,\,\,\,\Rightarrow\,\,\,\, z=-\frac{k_{1\dots i}\cdot\mathfrak{e}}{k_{1\dots i}\cdot\eta}\equiv z_{1\dots i}\,.\label{eq:zpole}
\end{equation}

\begin{figure}
\begin{centering}
\includegraphics[width=0.2\paperwidth]{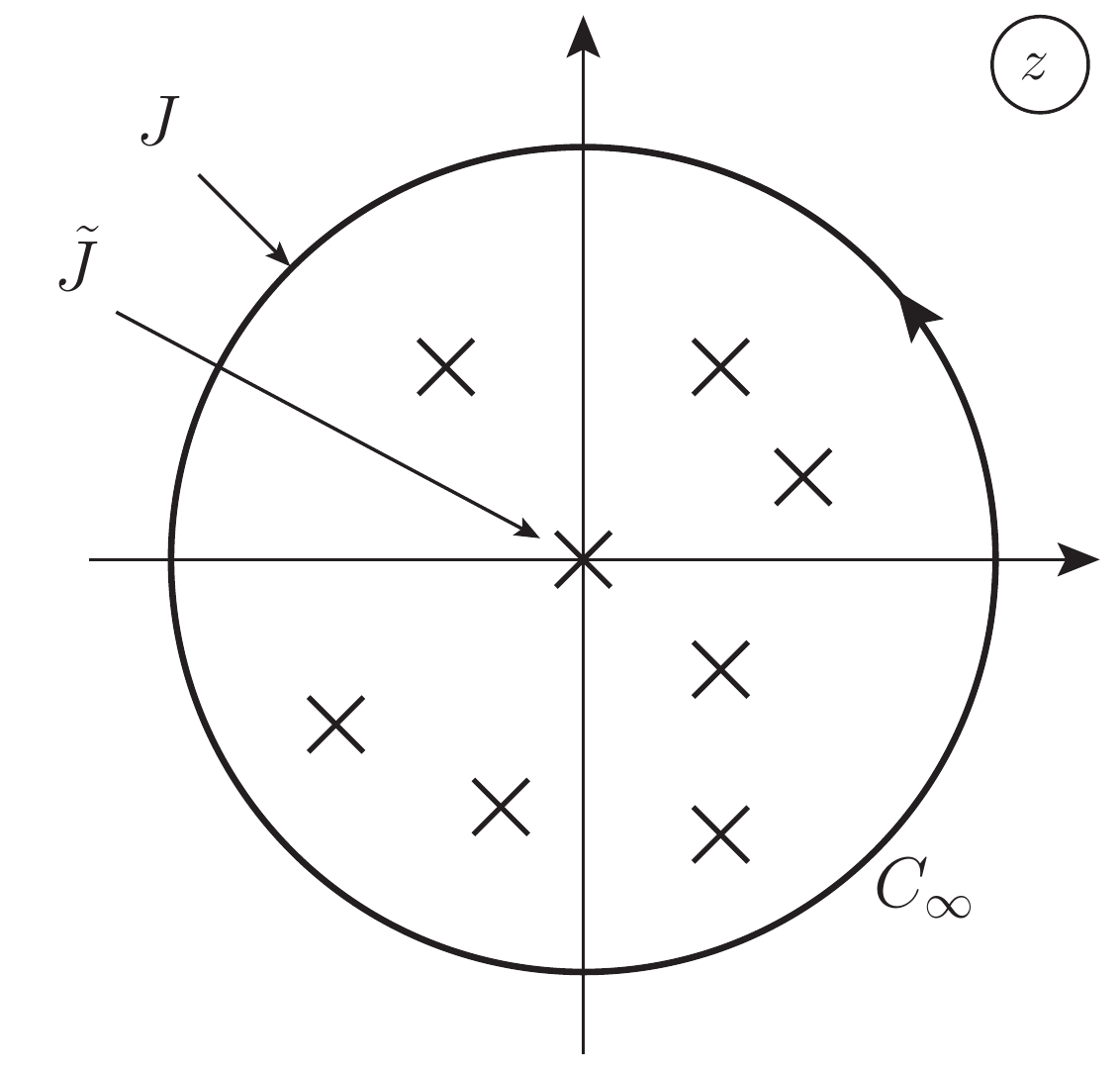}
\par\end{centering}
\caption{
Analytic structure of the matrix element of the straight infinite
Wilson line operator with the slope shifted by a complex four-vector $z\eta$. 
The residue at $z=0$ gives the gauge invariant off-shell current $\tilde{J}$, 
whereas the integral over the circle at infinity gives the off-shell current~$J$.
\label{fig:AnalyticStructure}
}
\end{figure}

Let us now consider an integral over $dz$ over a contour $C_{R}$
parametrized as $z=Re^{i\phi}$, $\phi\in[0,2\pi)$, $R\rightarrow\infty$,
enclosing all the poles of $f\left(z\right)$ (Fig.~\ref{fig:AnalyticStructure}).
The integral is non-zero and can be simply evaluated. 
As the dependence on $z$ appears only in the denominators and the first term of (\ref{eq:Mtildz1})
does not have any denominators, it is the only non-vanishing term for $\left|z\right|\rightarrow\infty$. 
Consequently the integral reads
\begin{equation}
\int_{C_{R}}dz\, f\left(z\right)=\,\mathfrak{e}\cdot J^{\left( \lambda_{1}\dots\lambda_{N}\right)}\left(k_{1\dots N}\right)=
J^{\, \left(\mathfrak{e}\rightarrow \lambda_{1}\dots\lambda_{N}\right)}\left(k_{1\dots N}\right).
\end{equation}

On the other hand, this integral is given by the sum of the residues
at the poles (\ref{eq:zpole}). In order to understand the structure
of a residue, say at some pole $z_{1\ldots i}$, let us refer to the
expansion (\arabic{Gaugelink_expansion}). We pick up the double-line
propagator giving rise to the pole (denoted by the shaded line in
the figure below) and observe that the terms can be resummed
into individual blobs to the left and to the right of the propagator; 
graphically:

\begin{flushleft}
\begin{tabular}{>{\centering}m{0.87\columnwidth}>{\centering}m{0.05\columnwidth}}
\medskip{}
\includegraphics[clip,height=0.06\paperheight]{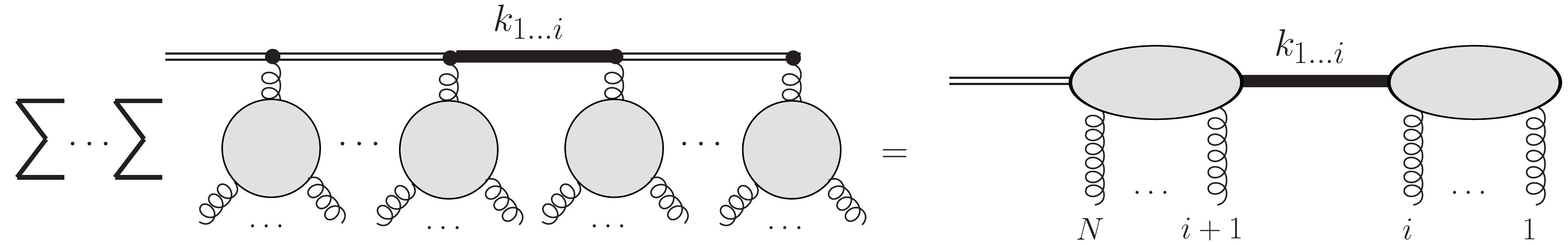} & \centering{}\centering{(\myref )}
\refmyref{WLresidue}\tabularnewline
\end{tabular}
\par\end{flushleft}

\noindent where the multiple sum on the left expresses all the contributions
containing the singular double-line propagator. The above identity
is most easily proved by expanding the r.h.s. using (\arabic{Gaugelink_expansion}) and comparing with the l.h.s. 
The blobs on the r.h.s correspond to $\tilde{J}$ with the Wilson line slope 
\begin{equation}
\mathfrak{e}_{1\dots i}=\mathfrak{e}+z_{1\dots i}\, \eta\,.
\end{equation}
Thus, the residue reads
\begin{eqnarray}
\mathrm{res}_{z_{1\dots i}}f\left(z\right) 
&=& 
i g \left[\frac{z-z_{1\dots i}}{z}\,
\tilde{J}^{\left(\mathfrak{e}(z)\rightarrow \lambda_{i+1}\dots\lambda_{N}\right)}\left(k_{(i+1)\dots N}\right)
\,\frac{-i}{k_{1\dots i}\cdot\left(\mathfrak{e}+z\eta\right)}\,
\tilde{J}^{ \left(\mathfrak{e}(z)\rightarrow \lambda_{1}\dots\lambda_{i}\right)}\left(k_{1\dots i}\right)\right]_{z=z_{1\dots i}} 
\nonumber \\
&=&
i g\, \tilde{J}^{\, \left(\mathfrak{e}_{1\dots i}\rightarrow \lambda_{i+1}\dots\lambda_{N}\right)}\left(k_{\left(i+1\right)\dots N}\right)\,\frac{i}{k_{1\dots i}\cdot\mathfrak{e}}\,
\tilde{J}^{\, \left(\mathfrak{e}_{1\dots i}\rightarrow \lambda_{1}\dots\lambda_{i}\right)}\left(k_{1\dots i}\right)\,.\label{eq:resid2}
\end{eqnarray}
Let us note that the rightmost $\tilde{J}$ is gauge invariant, as the condition $k_{1\dots i}\cdot\mathfrak{e}_{1\dots i}=0$ holds.
This is not the case for the leftmost $\tilde{J}$ current, however. 
The residue at $z=0$ is obviously the `unshifted' current
\begin{equation}
\mathrm{res}_{z=0}\, f\left(z\right)=\tilde{J}^{\, \left(\mathfrak{e}\rightarrow \lambda_{1}\dots\lambda_{N}\right)}\left(k_{1\dots N}\right).
\end{equation}

Having computed the residues and the boundary term, we can write the
final relation (Fig.~\ref{fig:AnalyticStructure}):
\begin{multline}
J^{\left(\mathfrak{e}\rightarrow \lambda_{1}\dots\lambda_{N}\right)}\left(k_{1\dots N}\right)
= \tilde{J}^{\, \left(\mathfrak{e}\rightarrow \lambda_{1}\dots\lambda_{N}\right)}\left(k_{1\dots N}\right)
\\
+ig \sum_{i=1}^{N-1}
\tilde{J}^{\, \left(\mathfrak{e}_{1\dots i}\rightarrow \lambda_{i+1}\dots\lambda_{N}\right)}\left(k_{\left(i+1\right)\dots N}\right)\,\frac{i}{k_{1\dots i}\cdot\mathfrak{e}}\,
\tilde{J}^{\, \left(\mathfrak{e}_{1\dots i}\rightarrow \lambda_{1}\dots\lambda_{i}\right)}\left(k_{1\dots i}\right) \, . 
\label{eq:MasterRel1}
\end{multline}
In respect of the general structure, this relation resembles the light-front Eq. (\ref{eq:RecRel1}). 
Indeed, as we shall see later on, when we specify the $\mathfrak{e}$ vector and the helicities, it reproduces the latter. 

The relation (\ref{eq:MasterRel1}) expresses the off-shell current $J^{\left(\mathfrak{e}\rightarrow \lambda_{1}\dots\lambda_{N}\right)}$
in terms of matrix elements of Wilson line operators 
and is  the central result of this paper.
The off-shell current emerged here as the boundary integral of the matrix element with shifted Wilson line slope.

In a similar manner we can construct the relation for the amplitude with an additional double-line propagator $\tilde{J}'$ defined in (\ref{eq:Mtildbar}). 
For this case, however, the integral over the contour $C_{\infty}$ vanishes.
The price we pay for the vanishing of the boundary term is that
now we have an additional pole coming from the additional leftmost propagator that we have included. 
Following the exact same procedure as before, we obtain the following relation
\begin{equation}
\tilde{J}'^{\, \left(\mathfrak{e}\rightarrow \lambda_{1}\dots\lambda_{N}\right)}\left(k_{1\dots N}\right)=ig\sum_{i=1}^{N}
\tilde{J}'^{\, \left(\mathfrak{e}_{1\dots i}\rightarrow \lambda_{i+1}\dots\lambda_{N}\right)}
\left(k_{\left(i+1\right)\dots N}\right)\,\frac{i}{k_{1\dots i}\cdot\mathfrak{e}}\,
\tilde{J}^{\, \left(\mathfrak{e}_{1\dots i}\rightarrow \lambda_{1}\dots\lambda_{i}\right)}
\left(k_{1\dots i}\right)\, ,
\label{eq:MasterRel2}
\end{equation}
with $\tilde{J}'\left(k_{\left(i+1\right)\dots N}\right)=\left(ig\right)^{-1}$ for $i=N$. 

We can provide an important interpretation of Eq. (\ref{eq:MasterRel2})
when we pull out the $N$-th residue explicitly and multiply the equation by $ik_{1\dots N}\cdot\mathfrak{e}$:
\begin{multline}
\tilde{J}^{\, \left(\mathfrak{e}\rightarrow \lambda_{1}\dots\lambda_{N}\right)}\left(k_{1\dots N}\right)=\tilde{J}^{\, \left(\mathfrak{e}_{1\dots i}\rightarrow \lambda_{1}\dots\lambda_{N}\right)}
\left(k_{1\dots N}\right)
\\
+ig\sum_{i=1}^{N-1}\tilde{J}^{\, \left(\mathfrak{e}_{1\dots i}\rightarrow \lambda_{i+1}\dots\lambda_{N}\right)}\left(k_{\left(i+1\right)\dots N}\right) \,
\frac{i k_{1\dots N}\cdot\mathfrak{e}}{k_{\left(i+1\right)\dots N}\cdot\mathfrak{e}_{1\dots i}\, 
k_{1\dots i}\cdot\mathfrak{e}}\,\tilde{J}^{\, \left(\mathfrak{e}_{1\dots i}\rightarrow \lambda_{1}\dots\lambda_{i}\right)}\left(k_{1\dots i}\right) \, . 
\label{eq:MasterRel2b}
\end{multline}
Essentially, this equation gives a prescription to change the Wilson
line slope from an arbitrary vector $\mathfrak{e}$ to a slope which
assures that $\tilde{J}$, the first term on the r.h.s., is gauge invariant.
A prescription of this type has been previously found for the MHV 
helicity configuration in \citep{Cruz-Santiago2015}, but it had a much more complicated form, with multiple sums.

Let us close this section by summarizing the essential features of and the differences between
the two relations derived here, Eqs. (\ref{eq:MasterRel1}) and (\ref{eq:MasterRel2b}):
\begin{enumerate}
\item the first relation (\ref{eq:MasterRel1}) expresses the current $\tilde{J}^{\, \left(\mathfrak{e}\rightarrow \lambda_{1}\dots\lambda_{N}\right)}$
as the off-shell current (boundary term) $J^{\left(\mathfrak{e}\rightarrow \lambda_{1}\dots\lambda_{N}\right)}$ plus a BCFW-type sum.
This holds irrespectively of gauge invariance, i.e. both whether $\mathfrak{e}\cdot k_{1\dots N} = 0$ or $\mathfrak{e}\cdot k_{1\dots N}\neq0$.
\item the second relation (\ref{eq:MasterRel2b}), instead, holds only for $\mathfrak{e}\cdot k_{1\dots N}\neq0$
and expresses $\tilde{J}^{\, \left(\mathfrak{e}\rightarrow \lambda_{1}\dots\lambda_{N}\right)}$
as the gauge invariant off-shell current $\tilde{J}^{\, \left(\mathfrak{e}_{1\dots N}\rightarrow \lambda_{1}\dots\lambda_{N}\right)}$
and the BCFW-type sum similar to the one in (\ref{eq:MasterRel1}), the only difference being in the intermediate propagator. 
\end{enumerate}
These relations can be applied in different situations, as we illustrate below.

\section{Applications}
\label{sec:Applications}

\subsection{The off-shell current $\left(+\rightarrow+\dots+\right)$}
\label{sub:plusplus}

As a first straightforward application of the new relations, 
we consider the off-shell current with helicity configuration $+\rightarrow+\dots+$.
We have already discussed this case in Section~\ref{sec:RecRel}
and the solution was given in Eq.~(\ref{eq:Mplusplus}). 
Here we will show how this solution can be simply obtained 
using new relations (\ref{eq:MasterRel2b}) and (\ref{eq:MasterRel1}).

First we shall solve (\ref{eq:MasterRel2b}) for a certain choice of $\mathfrak{e}$ 
and use that solution inside (\ref{eq:MasterRel1}) to find the final answer. To this end, let us define a polarization
vector for incoming `plus'-helicity gluon 
\begin{equation}
\varepsilon^{+\, *}\left(p\right)=\varepsilon^{-}\left(p\right)=\varepsilon_{\perp}^{-}+\frac{\vec{\varepsilon}_{\perp}^{\,\,-}\cdot\vec{p}_{\perp}}{p\cdot\eta}\eta \, ,
\end{equation}
for some auxiliary off-shell four-momentum $p\neq k_{1\dots N}$. 
Let us now substitute $\mathfrak{e}=\varepsilon^{-}\left(p\right)$ in (\ref{eq:MasterRel2b}).
As with this choice we have
\begin{equation}
\mathfrak{e}_{1\dots i}=\varepsilon^{-}\left(p\right)+z_{1\dots i}\eta=\varepsilon_{\perp}^{-} + 
\frac{\vec{\varepsilon}_{\perp}^{\,\,-}\cdot\vec{k}_{\perp1\dots i}}{k_{1\dots i}\cdot\eta}\eta=\varepsilon_{1\dots i}^{-} \, ,
\end{equation}
we get from (\ref{eq:MasterRel2b})
\begin{multline}
\tilde{J}^{\, \left(\varepsilon^{-}\left(p\right)\rightarrow +\dots +\right)}\left(k_{1\dots N}\right)=
\tilde{J}^{\, \left(\varepsilon^{-}_{1\dots N}\rightarrow  +\dots +\right)}\left(k_{1\dots N}\right)\\
+ig\sum_{i=1}^{N-1}
\tilde{J}^{\, \left(\varepsilon^{-}_{1\dots i}\rightarrow  +\dots +\right)}\left(k_{\left(i+1\right)\dots N}\right)\,\frac{ik_{1\dots N}\cdot\varepsilon^{-}\left(p\right)}{k_{\left(i+1\right)\dots N}\cdot\varepsilon_{1\dots i}^{-}\, k_{1\dots i}\cdot\varepsilon^{-}\left(p\right)}\,
\tilde{J}^{\, \left(\varepsilon^{-}_{1\dots i}\rightarrow  +\dots +\right)}\left(k_{1\dots i}\right)\,.\label{eq:Appl1}
\end{multline}
The gauge invariant off-shell current for the present helicity configuration
is zero, as proved in \citep{Cruz-Santiago2015},
\begin{equation}
\tilde{J}^{\, \left(\varepsilon^{-}_{1\dots i}\rightarrow  +\dots +\right)}\left(k_{1\dots i}\right)=
\tilde{J}^{\, \left( + \rightarrow  +\dots +\right)}\left(k_{1\dots i}\right) = 0,\,\,\,\textrm{for}\,\, i=2,\dots,N \, .
\label{eq:MplusGIzero1}
\end{equation}
Therefore we are left with only one non-vanishing term on the r.h.s of \eqref{eq:Appl1}. 
For $i=1$ we have
\begin{equation}
\tilde{J}^{\, \left(\varepsilon^{-}_{1}\rightarrow  +\right)}\left(k_{1}\right)=\,\varepsilon_{1}^{-}\cdot\varepsilon_{1}^{+}=-1\,.\label{eq:MplusGIzero2}
\end{equation}
Thus (\ref{eq:Appl1}) becomes
\begin{equation}
\tilde{J}^{\, \left(\varepsilon^{-}\left(p\right)\rightarrow +\dots +\right)}\left(k_{1\dots N}\right) = 
g\,\frac{\tilde{v}_{\left(1\dots N\right)\left(p\right)}}{\tilde{v}_{1\left(p\right)}\tilde{v}_{\left(2\dots N\right)1}} \,
\tilde{J}^{\, \left(\varepsilon^{-}_1\rightarrow +\dots +\right)}\left(k_{2\dots N}\right) \, ,
\label{eq:Appl2}
\end{equation}
where $\tilde{v}$ has been defined in (\ref{eq:vtild}). 
This is a simple recursion relation which can be readily solved by iteration.
Evaluating a few first terms \footnote{See the appendix of \citep{Cruz-Santiago2015}.},
it is easy to see that the candidate for a solution is
\begin{equation}
\tilde{J}^{\, \left(\varepsilon^{-}\left(p\right)\rightarrow +\dots +\right)}\left(k_{1\dots N}\right) = 
- \left(-g\right)^{N-1}\frac{\tilde{v}_{\left(1\dots N\right)\left(p\right)}}{\tilde{v}_{1\left(p\right)}}\,\frac{1}{\tilde{v}_{N\left(N-1\right)}\dots\tilde{v}_{32}\tilde{v}_{21}} \, .
\label{eq:Mplusgen1}
\end{equation}
The fact that the above expression is indeed the solution can be promptly checked
by inserting this into (\ref{eq:Appl2}). 
To this end, let us evaluate the $(N-2)$-th order object appearing on the r.h.s of (\ref{eq:Appl2}) using (\ref{eq:Mplusgen1}). We have 
\begin{equation}
\tilde{J}^{\, \left(\varepsilon^{-}_1\rightarrow +\dots +\right)}\left(k_{2\dots N}\right) = 
- \left(-g\right)^{N-2}\,\frac{\tilde{v}_{\left(2\dots N\right)1}}{\tilde{v}_{21}}\,\frac{1}{\tilde{v}_{N\left(N-1\right)}\dots\tilde{v}_{32}} \, .
\label{eq:Mplusgen2}
\end{equation}
Inserting this back into (\ref{eq:Appl2}) and noting that\footnote{See the appendix of \citep{Cruz-Santiago2015} also for many useful relations for $\tilde{v}$ quantities.}  $\tilde{v}_{\left(2\dots N\right)1}=\tilde{v}_{\left(1\dots N\right)1}$
we indeed obtain (\ref{eq:Mplusgen1}).
The expression in equation (\ref{eq:Mplusgen1}) is  not gauge-invariant -- and thus
non-zero with the present helicity configuration.

The off-shell current (\ref{eq:Mplusplus})
can be now easily recovered using (\ref{eq:MasterRel1}) with $\mathfrak{e}=\varepsilon_{1\dots N}^{-}$
and the expression (\ref{eq:Mplusgen2}). 
Using again (\ref{eq:MplusGIzero1}) and (\ref{eq:MplusGIzero2}) we get
\begin{equation}
J^{\,\left(+\rightarrow+\dots+\right)}\left(k_{1\dots N}\right)=g\,\frac{1}{\tilde{v}_{1\left(1\dots N\right)}}\tilde{J}^{\, \left(\varepsilon^{-}_1 \rightarrow +\dots +\right)}\left(k_{2\dots N}\right)=-\left(-g\right)^{N-1}\,\frac{\tilde{v}_{\left(1\dots N\right)1}}{\tilde{v}_{1\left(1\dots N\right)}}\frac{1}{\tilde{v}_{N\left(N-1\right)}\dots\tilde{v}_{32}\tilde{v}_{21}}\label{eq:Mplusplus1}
\end{equation}
which agrees with (\ref{eq:Mplusplus}). 
Note that in the two previous works \citep{Motyka2009} and \citep{Cruz-Santiago2015} 
this result was obtained with much more effort.

\subsection{The off-shell current $\left(+\rightarrow-+\dots+\right)$}
\label{sub:plusminus}

Let us now look at the next non-trivial off-shell current, that is the
MHV configuration. We set $\lambda_{1}=-$ and the remaining helicities
are all positive as before; remember that the incoming gluon
with $+$ helicity involves $\varepsilon_{1\dots N}^{-}$ polarization vector. 

For the present case it is not possible to derive the form of the gauge invariant off-shell current (\ref{eq:MHV})
from the relations (\ref{eq:MasterRel1}) or (\ref{eq:MasterRel2b}) alone,
as we shall explain in more detail in Section~\ref{sec:Discussion}.
However, we will show that the relation (\ref{eq:RecRel1}) obtained previously in \citep{Cruz-Santiago2013,Cruz-Santiago2015} 
emerges from (\ref{eq:MasterRel1}).

For the present helicity configuration the relation in Eq.~(\ref{eq:MasterRel1}) reads 
\begin{multline}
J^{\left(+ \rightarrow -+\dots+\right)}\left(k_{1\dots N}\right)=
\tilde{J}^{\,\left( +\rightarrow -+\dots +\right)}\left(k_{1\dots N}\right)\\
-ig\sum_{i=1}^{N-1}\tilde{J}^{\, \left(\varepsilon^{-}_{1\dots i}\rightarrow  +\dots +\right)}
\left(k_{\left(i+1\right)\dots N}\right)\,\frac{i}{\tilde{v}_{\left(1\dots i\right)\left(1\dots N\right)}}\,
\tilde{J}^{\, \left(+\rightarrow  -+\dots +\right)}\left(k_{1\dots i}\right)\,.\label{eq:MHVrel1}
\end{multline}
Above we have shortened the notation for the gauge invariant off-shell currents by writing
\begin{equation}
\tilde{J}^{\, \left(\varepsilon^{-}_{1\dots i} \rightarrow  -+\dots +\right)}\left(k_{1\dots i}\right)\equiv \tilde{J}^{\, \left(+\rightarrow  -+\dots +\right)}\left(k_{1\dots i}\right)\, .
\end{equation}

Using the result (\ref{eq:Mplusgen1}) let us write
\begin{multline}
\tilde{J}^{\, \left(\varepsilon^{-}_{1\dots i}\rightarrow  +\dots +\right)}\left(k_{\left(i+1\right)\dots N}\right)\,\frac{1}{\tilde{v}_{\left(1\dots i\right)\left(1\dots N\right)}}=
\frac{\left(-g\right)^{N-i-1}}{\tilde{v}_{\left(1\dots i\right)\left(1\dots N\right)}}\,\frac{\tilde{v}_{\left(i+1\dots N\right)
\left(1\dots i\right)}}{\tilde{v}_{\left(i+1\right)\left(1\dots i\right)}}\,\frac{1}{\tilde{v}_{N\left(N-1\right)}\dots\tilde{v}_{\left(i+2\right)\left(i+1\right)}}\\
= 
- K_{iN}\,J^{\left(+\rightarrow+\dots+\right)}\left(k_{\left(i+1\right)\dots N}\right),\label{eq:MHVrel2}
\end{multline}
where
\begin{equation}
K_{iN}=\frac{k_{1\dots N}^{+}}{k_{i+1\dots N}^{+}}\,\frac{1}{\tilde{v}_{\left(1\dots i+1\right)\left(i+1\right)}}\, , \label{eq:KiN}
\end{equation}
and $J$ is given in (\ref{eq:Mplusplus1}).
In order to obtain the above relation we have used some of the aforementioned properties of the $\tilde{v}$ quantities, in particular 
\begin{equation}
\tilde{v}_{ij}=-\frac{k_{i}^{+}}{k_{j}^{+}}\tilde{v}_{ji}\,.
\end{equation}
Plugging (\ref{eq:MHVrel2}) into (\ref{eq:MHVrel1}) and noticing that 
$\tilde{J}^{\, \left(\varepsilon_{1}^{-}\rightarrow - \right)}\left(k_{1}\right)=\tilde{J}^{\, \left(+\rightarrow - \right)}\left(k_{1}\right)=0$ we get 
\begin{multline}
\tilde{J}^{\,\left( +\rightarrow -+\dots +\right)}\left(k_{1\dots N}\right)=J^{\,\left(-\rightarrow+\dots+\right)}\left(k_{1\dots N}\right)\\
+g\sum_{i=2}^{N-1}\tilde{J}^{\, \left(+\rightarrow  -+\dots +\right)}\left(k_{1\dots i}\right)\, K_{iN}\,J^{\,\left(+\rightarrow+\dots+\right)}\left(k_{\left(i+1\right)\dots N}\right) \, ,
\end{multline}
which is precisely (\ref{eq:RecRel1}) -- the relation obtained in \citep{Cruz-Santiago2013,Cruz-Santiago2015},
after taking into account the different normalizations used here and in these references
(see footnote \footnotemark[\getrefnumber{footnote3}] ). 

\subsection{Explicit calculations of lower order currents}

In this subsection we present some explicit calculations of off-shell currents $J$ and their gauge invariant relatives $\tilde{J}$. 
The choice of the currents is such that they constitute building blocks which will allow to calculate the five-point off-shell next-to-MHV (NMHV) currents.

As we have already mentioned (see more extensive discussion in Section~\ref{sub:Noncloseness}), 
in general it is not possible to calculate $J$ or $\tilde{J}$ from our relations (\ref{eq:MasterRel1}) itself. 
In what follows we shall calculate the off-shell currents $J$ using the Berends-Giele recursion and then we will find $\tilde{J}$ from (\ref{eq:MasterRel1}). 
For the lower order examples presented below this is actually an over-kill as the same could be achieved by using directly  (\arabic{Gaugelink_expansion}). 
Nevertheless, this will provide a consistency check. Moreover, some of the results for $\tilde{J}$ presented here are new. 

Before we proceed, let us collect the triple and quartic gluon vertices in our notation as needed in the calculations. 
They are gathered in Table~\ref{tab:Triple-gluon-vertices} and Table~\ref{tab:Four-gluon-vertices}.

\begin{table}
\renewcommand{\arraystretch}{3.0}
\begin{centering}
\begin{tabular}{|c|c|}
\hline 
$V_{3}^{\left(+\rightarrow++\right)}\left(p,q\right)$ & $2ig\frac{k_{p+q}^{+}}{k_{p}^{+}}\,\tilde{v}_{pq}^{*}$
\tabularnewline
\hline 
$V_{3}^{\left(-\rightarrow--\right)}\left(p,q\right)$ & $2ig\frac{k_{p+q}^{+}}{k_{p}^{+}}\,\tilde{v}_{pq}$\tabularnewline
\hline 
$V_{3}^{\left(+\rightarrow-+\right)}\left(p,q\right)$ & $-2ig\frac{k_{q}^{+}}{k_{p+q}^{+}}\,\tilde{v}_{qp}$\tabularnewline
\hline 
$V_{3}^{\left(-\rightarrow-+\right)}\left(p,q\right)$ & $2ig\frac{k_{p}^{+}}{k_{p+q}^{+}}\,\tilde{v}_{pq}^{*}$\tabularnewline
\hline 
\end{tabular}\caption{Triple gluon vertices for some helicity configuration in the notation of the present paper.\label{tab:Triple-gluon-vertices}}

\par\end{centering}

\end{table}

\begin{table}
\centering{}%
\begin{tabular}{|c|c|}
\hline 
 &\\
$V_{4(\mathrm{LC})}^{\left(+\rightarrow-++\right)}\left(k,p,q\right)$ & $-ig^{2}\left(1-\frac{\left(2q^{+}+k^{+}+p^{+}\right)\left(k^{+}-p^{+}\right)}{\left(k^{+}+p^{+}\right)^{2}}\right)$\tabularnewline
&\\
\hline 
& \\
$V_{4(\mathrm{LC})}^{\left(-\rightarrow--+\right)}\left(k,p,q\right)$ & $-ig^{2}\left(1-\frac{\left(-q^{+}-p^{+}-2k^{+}\right)\left(p^{+}-q^{+}\right)}{\left(p^{+}+q^{+}\right)^{2}}\right)$\tabularnewline
& \\
\hline 
\end{tabular}\caption{Four-gluon vertices appearing in the light cone Berends-Giele relation for helicity configurations used in the paper.\label{tab:Four-gluon-vertices}}
\end{table}

\subsubsection{$\left(-\rightarrow--+\right)$}

Let us write (\ref{eq:MasterRel1}) for this specific case. We get simply
\begin{equation}
J^{\left(-\rightarrow--+\right)}\left(k_{123}\right)=\tilde{J}^{\left(\varepsilon_{123}^{+}\rightarrow--+\right)}\left(k_{123}\right)+g\tilde{J}^{\left(\varepsilon_{1}^{+}\rightarrow-+\right)}\left(k_{23}\right)\,\frac{1}{k_{1}\cdot\varepsilon_{123}^{+}}.\label{eq:Curr0}
\end{equation}

Let us calculate the off-shell current $J$ first. 
Actually, it is more convenient to calculate the off-shell amplitudes, i.e. to amputate the propagators. 
Thus we rewrite the formula above as
\begin{equation}
\mathcal{M}^{\left(-\rightarrow--+\right)}\left(k_{123}\right)=\tilde{\mathcal{M}}^{\left(\varepsilon_{123}^{+}\rightarrow--+\right)}\left(k_{123}\right)+g\,\frac{k_{123}^{2}}{k_{23}^{2}}\tilde{\mathcal{M}}^{\left(\varepsilon_{1}^{+}\rightarrow-+\right)}\left(k_{23}\right)\,\frac{1}{k_{1}\cdot\varepsilon_{123}^{+}}\,,\label{eq:amp1}
\end{equation}
where
\begin{equation}
\mathcal{M}\left(k\right)=ik^{2}J\left(k\right)\label{eq:MtoJrel} \, ,
\end{equation}
and similar for $\tilde{\mathcal{M}}$. Here $\tilde{\mathcal{M}}$
in the second term is just
\begin{equation}
\tilde{\mathcal{M}}^{\left(\varepsilon_{1}^{+}\rightarrow-+\right)}\left(k_{23}\right)=\varepsilon_{1\mu}^{+}d^{\mu\nu}\left(k_{23},\eta\right)V_{3\nu\alpha\beta}\left(k_{2},k_{3}\right)\varepsilon_{2}^{-\alpha}\varepsilon_{3}^{+\beta}=V_{3}^{\left(-\rightarrow-+\right)}\left(k_{2},k_{3}\right) \, .
\end{equation}

Let us calculate $\mathcal{M}^{\left(-\rightarrow--+\right)}\left(k_{123}\right)$.
There are four diagrams contributing
\begin{equation}
D_{1}=V_{3}^{\left(-\rightarrow-+\right)}\left(k_{12},k_{3}\right)\frac{-i}{k_{12}^{2}}\, V_{3}^{\left(-\rightarrow--\right)}\left(k_{1},k_{2}\right)\,,
\end{equation}
\begin{equation}
D_{2}=V_{3}^{\left(-\rightarrow-+\right)}\left(k_{1},k_{23}\right)\frac{-i}{k_{23}^{2}}\, V_{3}^{\left(+\rightarrow-+\right)}\left(k_{2},k_{3}\right)\,,
\end{equation}
\begin{equation}
D_{3}=V_{3}^{\left(-\rightarrow--\right)}\left(k_{1},k_{23}\right)\frac{-i}{k_{23}^{2}}\, V_{3}^{\left(-\rightarrow-+\right)}\left(k_{2},k_{3}\right)\,,
\end{equation}

\begin{equation}
D_{4}=V_{4(\mathrm{LC})}^{\left(-\rightarrow--+\right)}\left(k_{1},k_{2},k_{3}\right)\,.
\end{equation}
Using the expressions for the vertices and writing propagators as
\begin{equation}
k_{12}^{2}=-2\tilde{v}_{12}\tilde{v}_{21}^{*}\,,
\end{equation}
\begin{equation}
k_{23}^{2}=-2\tilde{v}_{23}\tilde{v}_{32}^{*}\,,
\end{equation}
we have
\begin{equation}
D_{1}=2i\frac{k_{12}^{+}}{k_{123}^{+}}\,\tilde{v}_{\left(12\right)3}^{*}\,\frac{-i}{-2\tilde{v}_{12}\tilde{v}_{21}^{*}}\,2i\frac{k_{12}^{+}}{k_{1}^{+}}\,\tilde{v}_{12}=-2ig^{2}\frac{\left(k_{12}^{+}\right)^{2}}{k_{1}^{+}k_{123}^{+}}\,\frac{\tilde{v}_{\left(12\right)3}^{*}}{\tilde{v}_{21}^{*}}\,,
\end{equation}
\begin{equation}
D_{2}=2i\frac{k_{1}^{+}}{k_{123}^{+}}\,\tilde{v}_{1\left(23\right)}^{*}\,\frac{-i}{-2\tilde{v}_{23}\tilde{v}_{32}^{*}}\,\left(-2i\right)\frac{k_{3}^{+}}{k_{23}^{+}}\,\tilde{v}_{32}=2ig^{2}\frac{k_{1}^{+}k_{3}^{+}}{k_{123}^{+}k_{23}^{+}}\,\frac{\tilde{v}_{1\left(23\right)}^{*}}{\tilde{v}_{23}^{*}}\,,
\end{equation}
\begin{equation}
D_{3}=2i\frac{k_{123}^{+}}{k_{1}^{+}}\,\tilde{v}_{1\left(23\right)}\,\frac{-i}{-2\tilde{v}_{32}\tilde{v}_{23}^{*}}\,2i\frac{k_{2}^{+}}{k_{23}^{+}}\,\tilde{v}_{23}^{*}=-2ig^{2}\frac{k_{123}^{+}k_{2}^{+}}{k_{1}^{+}k_{23}^{+}}\,\frac{\tilde{v}_{1\left(23\right)}}{\tilde{v}_{32}}\,,
\end{equation}
\begin{equation}
D_{4}=-ig^{2}\left[1+\frac{\left(k_{123}^{+}+k_{1}^{+}\right)\left(k_{2}^{+}-k_{3}^{+}\right)}{\left(k_{23}^{+}\right)^{2}}\right]\,.
\end{equation}
Adding the contributions we get nothing really illuminating. However,
if we calculate
\begin{equation}
\tilde{\mathcal{M}}^{\left(\varepsilon_{123}^{+}\rightarrow--+\right)}\left(k_{123}\right)=D_{1}+D_{2}+D_{3}+D_{4}+W,\label{eq:amp1-1}
\end{equation}
where $W$ is the gauge term on the r.h.s. of (\ref{eq:amp1})
\begin{equation}
W = 
-g\,\frac{k_{123}^{2}}{k_{23}^{2}}V_{3}^{\left(-\rightarrow-+\right)}\left(k_{2},k_{3}\right)\,\frac{1}{k_{1}\cdot\varepsilon_{123}^{+}}=
- 2ig^{2}\frac{\tilde{v}_{12}\tilde{v}_{21}^{*}+\tilde{v}_{13}\tilde{v}_{31}^{*} 
+ \tilde{v}_{23}\tilde{v}_{32}^{*}}{\tilde{v}_{23}\tilde{v}_{32}^{*}}\,\frac{k_{2}^{+}}{k_{23}^{+}}\,\frac{\tilde{v}_{23}^{*}}{\tilde{v}_{1\left(123\right)}^{*}}\,,
\end{equation}
we arrive at the following simple result
\begin{equation}
\tilde{\mathcal{M}}^{\left(\varepsilon_{123}^{+}\rightarrow--+\right)}
\left(k_{123}\right)=\tilde{\mathcal{M}}^{\left(-\rightarrow--+\right)}\left(k_{123}\right) = 
2ig^{2}\,\left(\frac{k_{3}^{+}}{k_{123}^{+}}\right)^{2}\,\frac{\tilde{v}_{\left(123\right)3}^{*3}}{\tilde{v}_{32}^{*}\tilde{v}_{21}^{*}\tilde{v}_{1\left(123\right)}^{*}}\,.\label{eq:Mtild---+}
\end{equation}

Let us note that this amplitude cannot be calculated as a `mostly plus' amplitude,
as a naive analogy with the on-shell case would suggest.
In that case, the result would read
\begin{equation}
2ig^{2}\,\left(\frac{k_{1}^{+}}{k_{2}^{+}}\right)^{2}\,\frac{\tilde{v}_{21}^{3}}{\tilde{v}_{\left(123\right)3}\tilde{v}_{32}\tilde{v}_{1\left(123\right)}} \, , 
\label{eq:Mtild---+_onshell}
\end{equation}
but (\ref{eq:Mtild---+}) and (\ref{eq:Mtild---+_onshell}) are not equal. 
They become equal in the on-shell limit $k_{123}^{2} \rightarrow 0$. 
Indeed, there are the following off-shell Schouten identities:
\begin{equation}
k_{123}^{2}=-2\left(\tilde{v}_{\left(123\right)3}\tilde{v}_{32}+\tilde{v}_{\left(123\right)1}\tilde{v}_{12}^{*}\right)\,,\label{eq:Sch1}
\end{equation}
\begin{equation}
k_{123}^{2}=-2\left(\tilde{v}_{\left(123\right)1}\tilde{v}_{13}+\tilde{v}_{\left(123\right)2}\tilde{v}_{23}^{*}\right)\,,\label{eq:Sch2}
\end{equation}
\begin{equation}
k_{123}^{2}=-2\left(\tilde{v}_{\left(123\right)2}\tilde{v}_{21}+\tilde{v}_{\left(123\right)3}\tilde{v}_{31}^{*}\right)\,.\label{eq:Sch3}
\end{equation}
When $k_{123}^{2}=0$, these are direct relations between relevant $\tilde{v}_{ij}$ (spinors in the on-shell case) and
(\ref{eq:Mtild---+}) becomes (\ref{eq:Mtild---+_onshell}), as expected.

The off-shell current can be now expressed via (\ref{eq:amp1}), (\ref{eq:MtoJrel}) in
a compact way, hardly visible when calculating directly from Feynman diagrams
\begin{equation}
J^{\left(-\rightarrow--+\right)}\left(k_{123}\right)=\frac{g^{2}}{\tilde{v}_{1\left(123\right)}^{*}}\Big[\frac{2}{k_{123}^{2}}\,\left(\frac{k_{3}^{+}}{k_{123}^{+}}\right)^{2}\,\frac{\tilde{v}_{\left(123\right)3}^{*3}}{\tilde{v}_{32}^{*}\tilde{v}_{21}^{*}}-\,\frac{1}{\tilde{v}_{32}}\frac{k_{2}^{+}}{k_{23}^{+}}\Big]\, . \label{eq:J---+}
\end{equation}

\subsubsection{$\left(+\rightarrow--+\right)$}

The relation (\ref{eq:MasterRel1}) adjusted to the present case reads
\begin{equation}
J^{\left(+\rightarrow--+\right)}\left(k_{123}\right)=\tilde{J}^{\left(\varepsilon_{123}^{-}\rightarrow--+\right)}
\left(k_{123}\right)\,.
\end{equation}
We see that that there are no `gauge-restoring' terms for this case. 
There are only two diagrams contributing to the amplitude and, for the off-shell current, 
the $k_{123}^2$ propagator has to be included according to Eq.~(\ref{eq:MtoJrel}).
\begin{equation}
D_{1}=V_{3}^{\left(+\rightarrow-+\right)}\left(k_{12},k_{3}\right)\frac{-i}{k_{12}^{2}}\, V_{3}^{\left(-\rightarrow--\right)}\left(k_{1},k_{2}\right)\,,
\end{equation}
\begin{equation}
D_{2}=V_{3}^{\left(+\rightarrow-+\right)}\left(k_{1},k_{23}\right)\frac{-i}{k_{23}^{2}}\, V_{3}^{\left(+\rightarrow-+\right)}\left(k_{2},k_{3}\right)\,.
\end{equation}
A similar calculation as before gives simply
\begin{equation}
J^{\left(+\rightarrow--+\right)}\left(k_{123}\right)=g^{2}\frac{k_{3}^{+}}{k_{123}^{+}}\,\frac{1}{\tilde{v}_{12}^{*}\tilde{v}_{23}^{*}} \, .
\end{equation}

\subsubsection{Five-point NMHV current $\left(+\rightarrow --++\right)$}

We are now ready to use the currents calculated above to obtain the five-point NMHV current. 
We start by calculating the off-shell current $J^{\left(+\rightarrow--++\right)}$
using the light-cone Berends-Giele recursion, which reads
\begin{eqnarray}
&&
ik_{1234}^{2}\, J^{\left(+\rightarrow--++\right)}\left(k_{1234}\right) = 
\nonumber \\ 
&& \hspace{3mm}
V_{3}^{\left(+\rightarrow-+\right)}\left(k_{1},k_{234}\right)J^{\left(+\rightarrow-++\right)}\left(k_{234}\right) + 
V_{3}^{\left(+\rightarrow-+\right)}\left(k_{12},k_{34}\right)\, J^{\left(-\rightarrow--\right)}\left(k_{12}\right)J^{\left(+\rightarrow++\right)}\left(k_{34}\right)
\nonumber \\
&&
+ V_{3}^{\left(+\rightarrow-+\right)}\left(k_{123},k_{4}\right)\, J^{\left(-\rightarrow--+\right)}\left(k_{123}\right)
+ V_{3}^{\left(+\rightarrow++\right)}\left(k_{123},k_{4}\right)\, J^{\left(+\rightarrow--+\right)}\left(k_{123}\right)
\nonumber \\
&&
+V_{4(\mathrm{LC})}^{\left(+\rightarrow-++\right)}\left(k_{12},k_{3},k_{4}\right)\, J^{\left(-\rightarrow--\right)}\left(k_{12}\right)
+V_{4(\mathrm{LC})}^{\left(+\rightarrow-++\right)}\left(k_{1},k_{23},k_{4}\right)\, J^{\left(+\rightarrow-+\right)}\left(k_{23}\right)\,.
\end{eqnarray}
Using the results from the previous subsections, we get explicitly
\begin{eqnarray}
&&
i^2k_{1234}^{2}\, J^{\left(+\rightarrow--++\right)}\left(k_{1234}\right) =
2\frac{k_{12}^{+}\left[k_{3}^{+}k_{1234}^{+}-k_{4}^{+}k_{12}^{+}\right]}{k_{123}^{+2}k_{2}^{+}}\,\frac{1}{\tilde{v}_{12}^{*}}
\nonumber \\
&&
+ 2\frac{k_{3}^{+} \left[k_{23}^{+}k_{1234}^{+}-k_{1}^{+}k_{4}^{+}\right]}{k_{123}^{+}k_{23}^{+}}\,\frac{1}{\tilde{v}_{23}^{*}} 
+ 2\frac{k_{12}^{+}k_{34}^{+2}}{k_{1234}^{+}k_{2}^{+}k_{4}^{+}}\,\frac{\tilde{v}_{\left(34\right)\left(12\right)}}{\tilde{v}_{34}\tilde{v}_{12}^{*}}
+ 2 \frac{k_{1234}^{+}k_{3}^{+}}{k_{123}^{+2}}\,\frac{\tilde{v}_{\left(123\right)4}^{*}}{\tilde{v}_{12}^{*}\tilde{v}_{23}^{*}} 
\nonumber \\
&&
+\frac{2k_{4}^{+}}{k_{1234}^{+}}\,\frac{\tilde{v}_{4\left(123\right)}}{k_{123}^{2}}\,\Bigg\{1+\frac{\left(k_{2}^{+}-k_{3}^{+}\right)\left(k_{1}^{+}+k_{123}^{+}\right)}{k_{23}^{+2}}
+\frac{2k_{2}^{+}k_{123}^{+}}{k_{1}^{+}k_{23}^{+}}\,\frac{\tilde{v}_{1\left(23\right)}}{\tilde{v}_{32}}+\frac{2k_{12}^{+}}{k_{123}^{+}}\,
\frac{\tilde{v}_{\left(12\right)3}^{*}}{\tilde{v}_{2\left(12\right)}^{*}}-\frac{2k_{1}^{+}k_{3}^{+}}{k_{23}^{+}k_{123}^{+}}\,\frac{\tilde{v}_{1\left(23\right)}^{*}}{\tilde{v}_{23}^{*}}\Bigg\}
\nonumber \\
&&
+ \frac{2k_{234}^{+}}{k_{1234}^{+}}\,\frac{\tilde{v}_{\left(234\right)1}}{k_{234}^{2}}\,\Bigg\{1-\frac{\left(k_{2}^{+}-k_{3}^{+}\right)\left(k_{4}^{+}+k_{234}^{+}\right)}{k_{23}^{+2}}
- \frac{2k_{2}^{+}k_{4}^{+}}{k_{23}^{+}k_{234}^{+}}\,\frac{\tilde{v}_{4\left(23\right)}}{\tilde{v}_{32}}+\frac{2k_{34}^{+2}}{k_{234}^{+}k_{4}^{+}} 
\, \frac{\tilde{v}_{\left(34\right)2}^{*}}{\tilde{v}_{34}^{*}}-\frac{2k_{3}^{+}k_{234}^{+}}{k_{23}^{+2}}\,\frac{\tilde{v}_{\left(23\right)4}^{*}}{\tilde{v}_{23}^{*}}\Bigg\}
\nonumber \\
\label{eq:J+--++}
\end{eqnarray}

This result can be greatly simplified when taking the on-shell limit.
In that case, the complex conjugate of the mostly minus MHV amplitude is obtained. 
However, this is not our goal. 
For the off-shell case, unfortunately, not much can be done,
due to `inhomogeneous' Schouten identities like those in (\ref{eq:Sch1})-(\ref{eq:Sch3}).

Having the off-shell current, we can use (\ref{eq:MasterRel1}) to compute its gauge invariant relative. 
Fixing the helicities to the NMHV configuration, we have
\begin{multline}
J^{\left(+\rightarrow--+\dots+\right)}\left(k_{1\dots N}\right) = 
\tilde{J}^{\left(\varepsilon_{1\dots N}^{-}\rightarrow--+\dots+\right)}\left(k_{1\dots N}\right) 
\\
-ig\sum_{i=3}^{N-1}\,\tilde{J}^{\left(\varepsilon_{1\dots i}^{-}\rightarrow--+\dots+\right)}\left(k_{\left(i+1\right)\dots N}\right)\,
\frac{i}{k_{1\dots i}\cdot\varepsilon_{1\dots N}^{-}}\,\tilde{J}^{\left(\varepsilon_{1\dots i}^{-}\rightarrow--+\dots+\right)}\left(k_{1\dots i}\right) \, .
\label{eq:NMHVany}
\end{multline}
The sum runs now from $i=3$ because
\begin{equation}
\tilde{J}^{\left(\varepsilon_{1\dots i}^{-}\rightarrow--+\dots+\right)}\left(k_{1\dots i}\right)=0,\,\,\, i=1,2\,.
\end{equation}

Let us now set $N=4$. We have
\begin{equation}
J^{\left(+\rightarrow--++\right)}\left(k_{1234}\right)=\tilde{J}^{\left(\varepsilon_{1234}^{-}\rightarrow--++\right)}\left(k_{1234}\right)-g\,\frac{1}{\tilde{v}_{\left(123\right)\left(1234\right)}}\, J^{\left(+\rightarrow--+\right)}\left(k_{123}\right)\,,\label{eq:NMHV5p}
\end{equation}
because
\begin{equation}
\tilde{J}^{\left(\varepsilon_{123}^{-}\rightarrow--+\right)}\left(k_{123}\right)=J^{\left(+\rightarrow--+\right)}\left(k_{123}\right)\,.
\end{equation}

Inserting the off-shell current (\ref{eq:J+--++}) into (\ref{eq:NMHV5p}) we obtain
\begin{eqnarray}
&&
\tilde{J}^{\left(\varepsilon_{1234}^{-}\rightarrow--++\right)}\left(k_{1234}\right)=\frac{-g^{3}}{k_{1234}^{2}}\Bigg\{2\frac{k_{12}^{+}\left[k_{3}^{+}k_{1234}^{+} 
- k_{4}^{+}k_{12}^{+}\right]}{k_{123}^{+2}k_{2}^{+}}\,\frac{1}{\tilde{v}_{12}^{*}} 
\nonumber \\
&&
+ 2 \frac{k_{3}^{+}\left[k_{23}^{+}k_{1234}^{+} - k_{1}^{+}k_{4}^{+}\right]}{k_{123}^{+}k_{23}^{+}}\,\frac{1}{\tilde{v}_{23}^{*}} 
+ 2\frac{k_{12}^{+}k_{34}^{+2}}{k_{1234}^{+}k_{2}^{+}k_{4}^{+}}\,\frac{\tilde{v}_{\left(34\right)\left(12\right)}}{\tilde{v}_{34}\tilde{v}_{12}^{*}}
+ 2\frac{k_{1234}^{+}k_{3}^{+}}{k_{123}^{+2}}\,\frac{\tilde{v}_{\left(123\right)4}^{*}}{\tilde{v}_{12}^{*}\tilde{v}_{23}^{*}}
\nonumber \\
&&
+\frac{2k_{4}^{+}}{k_{1234}^{+}}\,\frac{\tilde{v}_{4\left(123\right)}}{k_{123}^{2}}\,\Bigg[1+\frac{\left(k_{2}^{+}-k_{3}^{+}\right)\left(k_{1}^{+}+k_{123}^{+}\right)}{k_{23}^{+2}}
+\frac{2k_{2}^{+}k_{123}^{+}}{k_{1}^{+}k_{23}^{+}}\,\frac{\tilde{v}_{1\left(23\right)}}{\tilde{v}_{32}} 
+ \frac{2k_{12}^{+}}{k_{123}^{+}}\,\frac{\tilde{v}_{\left(12\right)3}^{*}}{\tilde{v}_{2\left(12\right)}^{*}} 
- \frac{2k_{1}^{+}k_{3}^{+}}{k_{23}^{+}k_{123}^{+}}\,\frac{\tilde{v}_{1\left(23\right)}^{*}}{\tilde{v}_{23}^{*}}\Bigg]
\nonumber \\
&&
+\frac{2k_{234}^{+}}{k_{1234}^{+}}\,\frac{\tilde{v}_{\left(234\right)1}}{k_{234}^{2}}\,\Bigg[1-\frac{\left(k_{2}^{+}-k_{3}^{+}\right)\left(k_{4}^{+}+k_{234}^{+}\right)}{k_{23}^{+2}}
-\frac{2k_{2}^{+}k_{4}^{+}}{k_{23}^{+}k_{234}^{+}}\,\frac{\tilde{v}_{4\left(23\right)}}{\tilde{v}_{32}} 
+\frac{2k_{34}^{+2}}{k_{234}^{+}k_{4}^{+}}\,\frac{\tilde{v}_{\left(34\right)2}^{*}}{\tilde{v}_{34}^{*}} 
- \frac{2k_{3}^{+}k_{234}^{+}}{k_{23}^{+2}}\,\frac{\tilde{v}_{\left(23\right)4}^{*}}{\tilde{v}_{23}^{*}}\Bigg]\Bigg\}
\nonumber \\
&&
+g^{3}\frac{1}{\tilde{v}_{\left(123\right)\left(1234\right)}}\,\frac{k_{3}^{+}}{k_{123}^{+}}\,\frac{1}{\tilde{v}_{12}^{*}\tilde{v}_{23}^{*}}  \, .
\label{eq:Jtild+--++}
\end{eqnarray}

Not much can be done with the above expression, even though it is gauge invariant. 
This is very different from the MHV configuration, where the inclusion of the `gauge-restoring' terms cancel the inhomogeneous terms in the Schouten identities 
and allows to obtain the result precisely in the form of the on-shell amplitude (see Eq.~(\ref{eq:MHV})). 
This confirms that the MHV off-shell amplitudes are special. 
Indeed, it appears from the CSW construction inspired by the twistor theory \citep{Cachazo2004}, that they are building blocks of any amplitude.

\section{Discussion}
\label{sec:Discussion}

\subsection{Beyond the MHV configuration}
\label{sub:BeyondMHV}

It is known that the MHV on-shell amplitudes are special, both because 
they are expressed by just one single term \citep{Parke:1986gb}, 
and they are building blocks of any on-shell amplitude via the CSW construction \citep{Cachazo2004}. 
Of course, in order to compute a complete scattering process for $N$ gluons, one needs
all helicity configurations, including next-to-MHV, next-to-next-to-MHV
and so on. These amplitudes will have a much more complicated structure than
the MHVs, even when the CSW construction is used. Similar statements are valid for gauge
invariant off-shell currents, as follows for example from our calculation presented in the previous section. 
Here the situation is even more complicated: as we have seen in our five-point NMHV example, 
the fact that we have an off-shell leg breaks the symmetry between the external legs. Thus the five-point NMHV current is not equal to the complex conjugate of the five-point MHV current with permuted legs, as this would be the case  in the on-shell case.

As discussed in Section~\ref{sec:RecRel}, the solution for the MHV
off-shell current given in the form (\ref{eq:RecRel1}) has been previously
obtained using the light-front quantization \citep{Cruz-Santiago2013}
and Wilson line technique \citep{Cruz-Santiago2015}. 
In both papers, however, the amount of algebra leading to the expression (\ref{eq:RecRel1})
was large enough to suggest that it was extremely unlikely to find a
similar expression for amplitudes beyond MHV level, mostly because  many 
algebraic manipulations were specific to the particular choice of the helicities.

However, our current analysis reveals that the  generalization of (\ref{eq:RecRel1}) 
beyond MHV-type helicity choice does exist and is equally simple in structure.
It is given in Eq.~(\ref{eq:MasterRel1}), which is valid in light-cone gauge for any helicity configuration. 
As we discuss below, unfortunately, this equation alone is not sufficient to calculate 
off-shell currents or gauge invariant amplitudes beyond the MHV order. It however reflects the general structure of the solution.

\subsection{Non-closeness of the new relations}
\label{sub:Noncloseness}

As we have already stated, the relation expressing off-shell currents
in terms of matrix elements of Wilson lines (\ref{eq:MasterRel1})
is in general not sufficient to calculate amplitudes themselves without an additional input. 
The basic reason is that they are not recursion
relations in  a strict sense because of the presence of the  boundary terms. 
That is, for an off-shell current having $N$ legs, the matrix element of the
Wilson line on the r.h.s. has also $N$ legs (the first term on the r.h.s. of (\ref{eq:MasterRel1})).
We shall refer to this property as non-closeness in what follows.

We have seen in Section~\ref{sub:plusplus}, how an additional information
regarding the gauge invariant off-shell current $\tilde{J}$ allowed to calculate
the off-shell current $J$ from (\ref{eq:MasterRel1})). Namely, for that case this external information
was that $\tilde{J}=0$ for the $+\rightarrow+\dots+$ helicity configuration. 
This case is however special and such property obviously does not hold for the other helicity configurations.

The non-closeness property can be more intuitively described in yet another way. 
Notice that the relation (\ref{eq:MasterRel1}) uses
only information about eikonal propagators  and couplings. 
It does not `know' anything about the internal structure of the interactions (i.e. triple or quartic gluon vertices) 
as is the case for the Berends-Giele recursion. 
This is in analogy to the soft gluon limit which does not allow to recover the full amplitude. 
Following this intuition, one may wonder why it does work for the $+\rightarrow+\dots+$ amplitude.
The reason is that, interestingly, for this case we have the following relation \citep{Cruz-Santiago2015}

\begin{flushleft}
\begin{tabular}{>{\centering}m{0.87\columnwidth}>{\centering}m{0.05\columnwidth}}
\medskip{}
\includegraphics[height=0.06\paperheight]{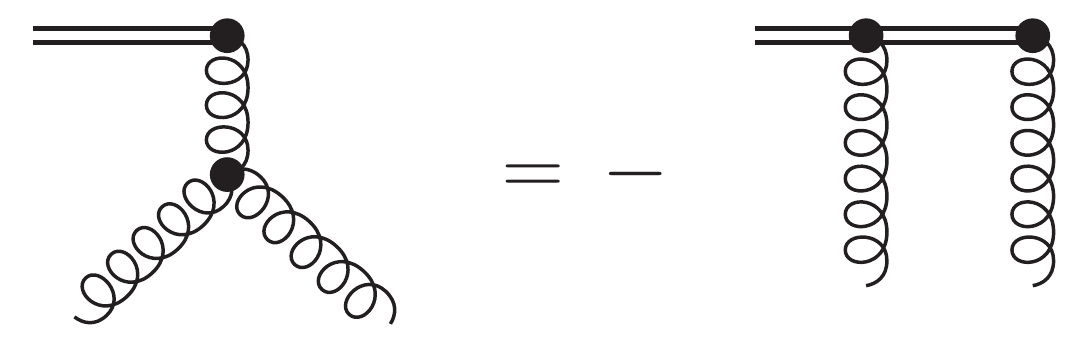} & \centering{}\centering{(\myref )}
\refmyref{eikV3}\tabularnewline
\end{tabular}
\par\end{flushleft}

\noindent which expresses the triple gluon vertex by the eikonal
diagram, allowing to calculate the amplitude using the relation (\ref{eq:MasterRel1}). 

To summarize, in order to calculate amplitudes from (\ref{eq:MasterRel1}) one needs to supplement it  with an
additional information and thus to close the system of equations. 
Let us note that the second equation can be always provided by the Berends-Giele recursion itself, i.e. by (\ref{eq:BerendsGiele2});
that is, one may successively trade off-shell currents in  (\ref{eq:BerendsGiele2}) for Wilson line matrix elements, and thus obtain a recursion.
However, this task is difficult to accomplish in general and is beyond the scope of the  present work.

\section{Summary and outlook}
\label{sec:Summary}

In this work we have investigated tree-level off-shell scattering amplitudes in pure Yang-Mills theory. 
More precisely, we have considered off-shell currents contracted with suitable polarization vectors, which are the building
blocks for the scattering amplitudes. We have shown that these off-shell
currents can be expressed in terms of objects which can be interpreted
as  matrix elements of Wilson line operators along a straight infinite path,
with the slope of the path corresponding to a certain polarization vector.
This construction emerges as a result of an integral over a complex shift of the Wilson
line slope. More precisely, in the light-cone gauge, we shift the
Wilson line slope in the direction of the light-cone gauge vector.
Such a shift allows to analyze the singularities of the Wilson line
propagators in the defined complex plane and derive certain relations in the spirit of the BCFW recursion relation. 
This result is given in Eq.~(\ref{eq:MasterRel1}). 
The construction itself is helicity independent and thus we were able to generalize a similar relation obtained
previously in \citep{Cruz-Santiago2013,Cruz-Santiago2015} for the case of the MHV configuration. 
Roughly half of the objects emerging in the new relation (\ref{eq:MasterRel1}) is gauge invariant, 
which accounts for the general simplicity of this equation.

Another interesting result is given in Eq.~(\ref{eq:MasterRel2b}), 
which relates Wilson line matrix elements with two different slopes. 
This relation generalizes and simplifies a similar relation 
found previously in \citep{Cruz-Santiago2015}, where 
it was used for several technical steps in formal derivations.

The derived relations are not closed, in the sense that they do not allow
to compute amplitudes unless additional input is given, as discussed in Section~\ref{sub:Noncloseness}. 
However, they  allow to trade an off-shell amplitude for matrix elements of Wilson lines.

Since in the present work we have studied a connection of Yang-Mills amplitudes and matrix elements of Wilson line operators, 
it is in order to mention the existing relations between QCD amplitudes and Wilson loops. 
First, there is a well known connection of this type realized by the Makeenko-Migdal approach \citep{Makeenko:1980vm} 
(an example  application to meson scattering  in the large $N_c$ limit is given in \citep{Makeenko:2011dm}).
Next, there is a famous duality between the Wilson loops in the 4-dimensional Super Yang-Mills theory 
and the on-shell scattering amplitudes  (see \cite{Brandhuber:2007yx} and  also \cite{Eden2013c,Eden2013b} for a review). This duality is very powerful and allows for calculating multi-loop correction (see for example \citep{Dixon:2015iva}). 
It appears though that the connection between the Wilson lines and the off-shell scattering amplitudes explored in the present work has a completely 
different sense than the aforementioned duality. 
Namely, the Wilson loops are used there to compute perturbative corrections to on-shell amplitudes, 
but the spinor structure itself is not related to the Wilson loops. Here, the Wilson line operators are used for gauge invariance reasons and are sensitive to the external polarizations.

It is interesting that the obtained relation (\ref{eq:MasterRel1}) is valid directly in the light-front quantization approach. 
This is because the Wilson line propagators do not involve the `minus' components. Thus the energy denominators and generally all light-front features would be hidden 
in the $J$, $\tilde{J}$ currents in~(\ref{eq:MasterRel1}). As showed in \citep{Cruz-Santiago2013} for the MHV case, on the light-front the sum of diagrams naturally collapses 
into the structure of (\ref{eq:MasterRel1}). Since we have proved that (\ref{eq:MasterRel1}) is valid for any helicity configuration, we may expect that similar natural 
resummation of diagrams should occur in any amplitude calculation on the light-front. It would be interesting to demonstrate this statement explicitly, 
although this would be technically very demanding.

Numerous further avenues could be explored as possible applications of the results presented in this work. 
One possibility is the improvement of the QCD evolution equations for unintegrated gluon densities at high energy, by incorporating more accurate kinematics.
The  famous 
Balitsky-Fadin-Kuraev-Lipatov (BFKL) \cite{Fadin:1975cb,Balitsky:1978ic} evolution equation for the unintegrated parton density and its nonlinear extensions, 
the Balitsky-Kovchegov (BK) and Jalilian-Marian-Iancu-McLerran-Weigert-Leonidov-Kovner (JIMWLK) equations \cite{Balitsky1996,Kovchegov:1999yj,JalilianMarian:1997gr,Iancu:2001ad}, 
are currently known  up to next-to-leading logarithmic order \cite{Fadin:1998py,Balitsky:2008zza,Kovner:2014lca}.
These equations resum the cascade of multiple gluon emissions in QCD in the high energy limit.
It is known that the next-to-leading corrections to these evolution equations are numerically large and they mostly originate from the kinematics. 
Resummation schemes have been proposed \cite{Ciafaloni:2003rd}, which contain more exact treatment of the kinematics along the cascade which goes beyond the high energy limit.  
It would be desirable to improve upon these approximate resummation schemes  and obtain the  description of the gluon cascade 
which would include complete kinematic information along the cascade. The information about the
general structure of the off-shell multi-gluon amplitudes could be very important in this case (for preliminary work in this direction see \cite{Motyka2009}).
The relations for the off-shell amplitudes derived in this paper might be useful in constructing such improved evolution equations.

Another interesting line of research would be to 
generalize the BCFW recursion obtained in \citep{vanHameren:2014iua,vanHameren:2015bba} for off-shell high-energy amplitudes used in $k_T$ factorization.
In terms of Wilson lines, these amplitudes correspond to the Wilson line slope given by the light-like component of the off-shell momentum while in the present work the slope was set to the transverse polarization vectors.
We think that an extension of \citep{vanHameren:2014iua,vanHameren:2015bba} to account for arbitrary slope would be very desirable and would open new possibilities in the context of the present paper.

\section*{Acknowledgments}
This work was supported by the Department of Energy  Grants  No. DE-SC-0002145, DE-FG02-93ER40771, by
the National Science Center, Poland, Grant No. 2015/17/B/ST2/01838 and by the Angelo Della Riccia foundation. 
MS wishes to thank for hospitality the Penn State University where part of this project was developed. 
The authors would like to thank Leszek Motyka and Andreas van~Hameren for useful discussions. 

\bibliographystyle{JHEP}
\bibliography{library}

\end{document}